\documentclass[11pt,a4paper]{article}
\pdfoutput=1
\usepackage{jheppub}
\usepackage{slashed}

\newcommand{\slv}{\raise.15ex\hbox{$/$}\kern-.53em\hbox{$v$}}
\newcommand{\slF}{\raise.15ex\hbox{$/$}\kern-.53em\hbox{$F$}}
\newcommand{\slL}{\raise.15ex\hbox{$/$}\kern-.53em\hbox{$L$}}
\newcommand{\slP}{\raise.15ex\hbox{$/$}\kern-.53em\hbox{$P$}}
\newcommand{\slp}{\raise.15ex\hbox{$/$}\kern-.53em\hbox{$p$}}
\newcommand{\slq}{\raise.15ex\hbox{$/$}\kern-.53em\hbox{$q$}}
\newcommand{\slR}{\raise.15ex\hbox{$/$}\kern-.53em\hbox{$R$}}
\newcommand{\slQ}{\raise.15ex\hbox{$/$}\kern-.53em\hbox{$Q$}}
\newcommand{\slK}{\raise.15ex\hbox{$/$}\kern-.53em\hbox{$K$}}
\newcommand{\slk}{\raise.15ex\hbox{$/$}\kern-.53em\hbox{$k$}}
\newcommand{\slD}{\raise.15ex\hbox{$/$}\kern-.53em\hbox{$D$}}
\newcommand{\slC}{\raise.15ex\hbox{$/$}\kern-.53em\hbox{$C$}}
\newcommand{\slA}{\raise.15ex\hbox{$/$}\kern-.53em\hbox{$A$}}
\newcommand{\slSigma}{\raise.15ex\hbox{$/$}\kern-.53em\hbox{$\Sigma$}}
\newcommand{\slpartial}{\raise.15ex\hbox{$/$}\kern-.53em\hbox{$\partial$}}
\newcommand{\slcalP}{\raise.15ex\hbox{$/$}\kern-.63em\hbox{$\cal P$}}

\def\bs{\boldsymbol}
\def\qqb{{q\bar q}}
\def\q{{\boldsymbol q}}
\def\kh{{\boldsymbol \kappa}}
\def\khb{\bar{{\boldsymbol \kappa}}}
\def\Qh{\kh - \q}
\def\Qhb{\khb - \q}

\def\pip{\delta{\boldsymbol n}}
\def\Omegaq{\frac{(\Qh)^2}{2k^+}}
\def\Omegaqzer{\frac{\kh^2}{2k^+}}
\def\Omegaqb{\frac{(\Qhb)^2}{2k^+}}
\def\Omegaqbzer{\frac{\khb^2}{2k^+}}
\def\Omegaqqb{\Xi_\qqb}

\def\pb{\bar p}
\def\pp{{\boldsymbol p}}
\def\ppb{{\bar {\boldsymbol p}}}

\def\epsp{{\boldsymbol \epsilon}_\lambda}

\def\x{{\boldsymbol x}}

\def\r{{\boldsymbol r}_\perp}

\def\Q{{(\boldsymbol {k-q})}}

\def\bdel{{\boldsymbol \del}_\perp}

\def\Rq{\mathcal{R}_q}
\def\Rqb{\mathcal{R}_{\bar q}}
\def\Rcoh{\mathcal{R}_\text{sing}}
\def\J{\mathcal{J}}

\def\Pq{\mathcal{P}_q}
\def\Pqb{\mathcal{P}_{\bar q}}
\def\sM{\text{med}}
\def\A{\mathcal{A}}
\def\aabb{\Omega_\text{a}}
\def\abab{\Omega_\text{b}}
\def\trdiff{\Omega_\text{ab}}
\def\potint{\int_\mathcal{V(\q)}}

\def\emcurr{{\bs C}}

\newcommand{\beq}{\begin{eqnarray}}
\newcommand{\eeq}{\end{eqnarray}}

\long\def\comment#1{ }    
\newcommand{\be}{\begin{equation}}
\newcommand{\ee}{\end{equation}}

\newcommand{\nn}{\nonumber\\ }

\newcommand{\labe}{\label}
\def\del{\partial}

\renewcommand{\tabcolsep}{0.5cm}

\title{The radiation pattern of a QCD antenna in a dilute medium}

\author[a,b]{Yacine~Mehtar-Tani,}

\author[b]{Carlos~A.~Salgado}

\author[c]{and Konrad~Tywoniuk}

\affiliation[a]{Institut de Physique Th\'eorique, CEA Saclay, 
F-91191 Gif-sur-Yvette, France}

\affiliation[b]{Departamento de F\'isica de Part\'iculas, 
Universidade de Santiago de Compostela, \\
E-15782 Santiago de Compostela, 
Galicia-Spain}

\affiliation[c]{Department of Astronomy and Theoretical Physics, 
Lund University, S\"olvegatan 14A,
SE-223 62 Lund, 
Sweden}

\emailAdd{yacine.mehtar-tani@cea.fr}
\emailAdd{carlos.salgado@usc.es}
\emailAdd{konrad.tywoniuk@thep.lu.se}

\abstract{

Radiative interferences in the multi-parton shower is the building block of QCD jet physics in vacuum. The presence of a hot medium made of quarks and gluons is expected to alter this interference pattern. To study these effects, we derive the gluon emission spectrum off a color-correlated quark-antiquark pair (antenna) traversing a colored medium to first order in the medium density. The resulting induced gluon distribution is found to be governed by the hardest scale of the problem. In our setup, this can either be the inverse antenna transverse size, $r_\perp^{-1}$, or the scale related to the transverse color correlation length in the medium, which is given by the Debye mass $m_D$. This emerging scale opens the angular phase space of emissions off the antenna compared to the vacuum case and gives rise to a typical transverse momentum of the medium-induced emitted gluons, $\langle k_\perp^2 \rangle_{\text{med}}\sim \text{max}(r_\perp^{-1},m_D)$. Above the hard scale interference effects suppress the spectrum resulting in the restoration of vacuum coherence.

}


\arxivnumber{1112.5031}
\dedicated{Preprint: LU-TP 11-18}

\begin{document}

\maketitle

\section{Introduction}
\label{sec:intro}

The recent start-up of the highly versatile experimental program at the Large Hadron Collider (LHC) at CERN for collisions of both protons and nuclei motivates a closer look at medium effects on hard probes. For heavy-ion collisions, in particular, considerable interest is associated with the measurement of the properties of the final state fragmentation processes of energetic quarks and gluons traversing the dense region created in the aftermath of the collision.
Results on single-particle inclusive spectra \citep{Aamodt:2010jd} and di-jet energy asymmetry \citep{Aad:2010bu,Chatrchyan:2011} in Pb+Pb collisions at $\sqrt{ s_{NN}} = 2.76$ TeV reveal strong effects with respect to the p+p baseline. These initial results, together with the first jet studies in Au+Au collisions at $\sqrt{s_{NN}}=200$ GeV performed at RHIC \citep{Putschke:2008wn,Salur:2008hs,Lai:2009zq}, mark the beginning of a new era for perturbative physics in heavy-ion collisions. 

While the physical mechanisms of fragmentation in vacuum are well-known and controlled to high accuracy, in this work we will mainly be concerned with the modifications arising due to the presence of a deconfined medium. First and foremost, one expects the spatiotemporal evolution of the medium to interfere with the relevant time-scales of emissions. The presence of a spatially extended region of color charges could hence affect the radiation pattern of emitted gluons, in particular, disturbing the subtle interference effects between emitters. These issues were first addressed in the context of gluon radiation off a quark--antiquark ($\qqb$) antenna in refs.~\citep{MehtarTani:2010ma,MehtarTani:2011tz}. 

Due to the soft and collinear divergences accompanying gluon emissions off an energetic projectile in vacuum, the smallness of the strong coupling constant can be compensated by large logarithms of the available phase space. This requires a resummation of multiple emissions and leads to the picture of a parton cascade. The appearance of multiple emitters, in turn, gives rise to color coherence effects which impose strict angular ordering of subsequent emissions \citep{Dokshitzer:1991wu,Mueller:1981ex,Ermolaev:1981cm}. Put more generally, soft gluon radiation is only sensitive to the total charge of the emitting system. These properties establish the factorization of subsequent emissions which lies at the heart of the jet calculus \citep{bas83,Konishi:1979cb}. The semi-classical, Markovian nature of the process also makes possible an implementation using Monte-Carlo simulations. 

In contrast to the vacuum case, the medium-induced radiative gluon spectrum off a single emitter contains neither infrared nor collinear divergences \citep{Baier:1996kr,Baier:1996sk,Zakharov:1996fv,Zakharov:1997uu,Wiedemann:2000ez,Wiedemann:2000za,gyu00}. Moreover, gluons produced with formation times larger than the spatial extension of the medium are strongly suppressed. The resulting distribution gives  a dominant contribution at a characteristic gluon energy and emission angle determined by local medium properties. Thus, a dense medium induces a strong depletion of the energy of the leading particle due to the radiative process. This energy loss of hard particles inside the medium has been extensively studied experimentally at RHIC by measuring the suppression of particles produced at high transverse momentum \citep{RHIC,Back:2004je,Arsene:2004fa,Adams:2005dq}.

One of the main limitations of these formalisms of jet quenching is the handling of multiple emissions. Since, by construction, the independent spectrum off the quark does not incorporate effects of interference among several emitters, which are fundamental for building up of the shower picture in vacuum, there is no {\it a priori} procedure of extending these results. In the literature, several approaches have been suggested, see refs.~\citep{CasalderreySolana:2007zz,Majumder:2010qh,Armesto:2011ht} for comprehensive reviews. Here, we briefly mention two of the most typical approaches.  On one hand, one can assume the multiple emissions to be independent of each other \citep{Baier:2001yt,sal03}. This implies a Poisson distribution of multiple medium-induced gluon emissions. Such a procedure explicitly ignores any correlation between various emissions. On the other hand, one have incorporated the medium-induced spectrum as a modification of the standard Altarelli-Parisi splitting function in vacuum \citep{Guo:2000nz,Majumder:2004pt,Armesto:2007dt}. This implicitly assumes the same ordering variables for the medium-induced radiation as for the vacuum one. See also refs.~\citep{Majumder:2010qh,Armesto:2011ht} for other proposals.

Ultimately, the approaches described above are only heuristically motivated and provide working hypotheses for phenomenological applications. In order to establish a consistent showering picture and possibly identify the corresponding ordering variable for subsequent emission, an analysis of the interferences arising between various emitters is essential. The purpose of this work is to overcome some of the limitations mentioned above, in particular studying the case of gluon emission off two emitters -- a quark-antiquark pair formed initially in either a color singlet or a color octet state. This setup allows to address the color coherence effects analogous to those responsible for angular ordering in vacuum. 

The present paper provides a detailed discussion on the mathematical derivation and the results first presented in ref.~\citep{MehtarTani:2010ma}. Besides, we generalize the calculations to arbitrary color representation of the parent parton, i.e., in our case $\gamma^\ast\to\qqb$ (singlet) and $g^\ast\to\qqb$ (octet). As in ref.~\citep{MehtarTani:2010ma}, we will for the time being restrict our calculation to first order in the medium density without losing generality, in the sense that all relevant transverse interference effects are contained in the calculation. The resummation of all orders in medium opacity, which improves the treatment of longitudinal interference among scattering centers, was presented in ref. \citep{MehtarTani:2011tz} for gluon emissions in the soft limit and extended to finite gluon energy in refs. \citep{MehtarTani:2011jw,CasalderreySolana:2011rz}. We work  in the framework of classical Yang-Mills (CYM) equations which is appropriate for soft gluon production, e.g., see ref.~\citep{MehtarTani:2006xq}. Nevertheless, for the sake of a proper diagrammatic interpretation of our result, we have also calculated, in parallel, all relevant Feynman diagrams. The details of the latter calculation can be found in Appendix~\ref{sec:feynman}. A numerical analysis of the antenna spectrum, including the case of massive quarks, have already been presented in ref.~\citep{arXiv:1110.4343}.

The medium-induced coherent spectrum differs significantly from the independent one \citep{Baier:1996kr,Baier:1996sk,Zakharov:1996fv,Zakharov:1997uu,Wiedemann:2000ez,Wiedemann:2000za,gyu00} due to the inclusion of interference diagrams which were heretofore neglected. These new contributions die out only in the case of very hard gluon emissions and for large opening angles of the pair. For observables that are biased towards such configurations, e.g., the energy loss of a leading jet particle, interferences have a small impact. On the other hand, for typical circumstances relevant for parton cascades in heavy-ion collisions these novel contributions are expected to alter the soft structure of the jet. Specifically, in the soft limit the cross section is logarithmically enhanced due to the soft divergence and exhibits a vacuum-like out-of-cone radiation pattern  \citep{MehtarTani:2010ma}.

It will be shown that the antenna bulk spectrum in general relies only on two characteristic transverse scales which have a straightforward geometrical interpretation, see Section \ref{sec:scaling} for further details. The first relevant scale is given by the size of the antenna-dipole probed by the medium, $r_\perp = \theta_\qqb L$, where $\theta_\qqb$ is the opening angle and $L$ is the length of the medium. The second one is related to the transverse color correlation length of the medium, $m_D^{-1}$, the inverse of the Debye mass parameter of the Yukawa-type medium interaction potential.\footnote{Interference effects between scattering centers at different longitudinal positions modifies this estimate by a factor $\sqrt{L/\lambda}$, where $\lambda$ denotes the mean free path of the medium.} When the antenna probes the medium at transverse distances smaller than this correlation length, $r_\perp \ll m_D^{-1}$, it interacts as a whole and, thus, preserves the coherent character of the emissions. In this case the medium opens the angular phase space for bremsstrahlung emissions. The strength of the interaction is controlled by the decoherence parameter, $\Delta_\text{med} \propto r_\perp^2$, which reveals the typical feature of color transparency. Since the spectrum in this situation is completely given by the characteristics of the $\qqb$ dipole, we call it the ``dipole" regime. In the opposite case, $r_\perp \gg m_D^{-1}$, the medium is predominantly probing the inner structure of the antenna. Coherent emissions can in this case only take place up to momentum scales $\sim m_D$ at which point radiation is mainly induced independently off each of the constituents. Since the decoherence parameter is saturated at its maximal value in this regime so that the size of the $\qqb$ dipole cease to play a role, we denote it as the ``saturation" regime. 

Thus, the hard scale which governs the main features of the spectrum is given by $Q_\text{hard} \equiv \max \big(r_\perp^{-1}, m_D \big)$. This scale roughly dictates the maximum transverse momentum $k_\perp$ that can be transferred to the induced gluon. In both regimes the resulting gluon spectrum is leading for $k_\perp \lesssim Q_\text{hard}$ and strongly suppressed for larger $k_\perp$. The general picture is  summarized in Table~\ref{tab:summary}. 
\begin{table}[t]
\center
\begin{tabular}{l || c | l}
 & $Q_\text{hard}$ & regime \\
 \hline
 $r_\perp \ll m_D^{-1}$ & $r_\perp^{-1}$ & ``dipole" \\
 $r_\perp \gg m_D^{-1}$ & $ m_D $ & ``saturation" 
 \end{tabular}
 \label{tab:summary}
 \caption{Two regimes of the bulk antenna spectrum.}
\end{table}

The outline of the paper is as follows. In Section \ref{sec:vacuum} we recall the key features of the radiation spectrum off a $\qqb$ antenna in vacuum, such as angular ordering, and present the CYM formalism and our notations. Section \ref{sec:medium} contains the calculation of the medium-indued spectrum off a $\qqb$ antenna in an initially color singlet state at first order in the background field. Our main result is given in eq.~(\ref{eq:SpectrumMed}). Among the contributions, we identify the well-known independent emission spectra off the quark and antiquark, as well as novel interference terms. Before inquiring into the novel features, at the outset we present a short review of the independent spectrum in Section \ref{sec:glv}.  The interferences contain a contribution which becomes dominant in the infrared limit. We analyze this term analytically in Section~\ref{sec:soft} and find that radiation is only allowed outside of the cone delimited by the $\qqb$ opening angle, a feature we call antiangular ordering. Here, we also identify the two regimes of induced radiation which determines the emerging scale dependence of the total spectrum beyond the soft limit. We discuss these aspects in detail in Section \ref{sec:scaling}. For soft gluons emitted in the ``saturation" regime, described in Section \ref{sec:saturationregime}, two distinct emission mechanisms are clearly separated, allowing for a probabilistic interpretation. This extends the physics of decoherence in medium to arbitrary gluon energies. A short summary is presented in Section \ref{sec:summary}. 

We define the coherent spectrum off the quark in Section \ref{sec:numerics} which also contains the numerical analysis of the antenna gluon spectrum and go on to discuss the general features of the spectrum found analytically in the preceding sections.  In particular, we demonstrate the role of the characteristic hard scale $Q_\text{hard}$ in the ``dipole" and ``saturation" regimes in Sections \ref{sec:numericsdipole} and \ref{sec:numericssaturation}, respectively. Finally, we demonstrate that the above features are also relevant for the color octet antenna in Section \ref{sec:octet}. We conclude and discuss the implications of the novel effects on jet physics in heavy-ion collisions in Section \ref{sec:conclusions}.

\section{Antenna radiation in vacuum}
\label{sec:vacuum}

Let us first review the vacuum emission pattern off a quark-antiquark ($\qqb$) pair which will serve as a short reminder of the basic steps involved and as a introduction to the  semi-classical calculation. In the classical limit, the inclusive spectrum for gluon radiation with momentum $k=(\omega,{\vec k})$ is given by 
\be
(2\pi)^3 2\omega \ \frac{dN}{d^3k}=\sum_{\lambda=\pm 1} {\cal M}_\lambda^{a}({\vec k}){\cal M}_\lambda^{\ast a}({\vec k}) \;,
\ee
and the amplitude is related to the classical gauge field by the reduction formula 
\be\label{eq:redform}
{\cal M}_\lambda ^{a}({\vec k})=\lim_{k^2\to 0 } -k^2A^{a}_\mu(k)  \epsilon^\mu_\lambda({\vec k}) \;,
\ee
where only the physical, transverse polarizations of the gluon contribute to the cross-section, so that $\sum_\lambda \epsilon_\lambda^i (\epsilon_\lambda^j)^\ast = \delta^{ij}$ ($i,j=$ 1,2). The gauge field, $A^\mu\equiv A^{\mu,a}t^a$, where $t^a$ is the generator of SU(3) in the fundamental representation, is the solution of the CYM equations,
\beq
\label{eq:CYMvacuum}
[D_\mu,F^{\mu\nu}]=J^\nu \,,
\eeq
where $D_\mu\equiv \del_\mu-ig A_\mu$ is the covariant derivative and $F_{\mu\nu}\equiv \del_\mu A_\nu-\del_\nu A_\mu-ig[A_\mu,A_\nu]$ is the non-Abelian field strength tensor. The current $J^\mu$, which describes the projectiles, is covariantly conserved, i.e., $[D_\mu,J^\mu]=0$. In the following, we will work in the physical, light-cone (LC) gauge, $A^+=0$.\footnote{The light-cone decomposition of the 4-vector $x\equiv(x^0,x^1,x^2,x^3)$ is defined as $x\equiv(x^+,x^-,\x)$, where $x^\pm\equiv(x^0 \pm x^3)/\sqrt{2}$ and $\x = (x^1,x^2)$.} Then, due to the gauge condition $ \epsilon^+_\lambda(k) = 0$, the transversality $k\cdot \epsilon_\lambda(k) = 0$, leads to the LC decomposition of the polarization vector $\epsilon_\lambda = (0,\frac{{\bs k} \cdot \epsp}{ k^+},\epsp)$.

\begin{figure}\label{fig:AntennaVacuum}
\centering
\includegraphics[width=0.7\textwidth]{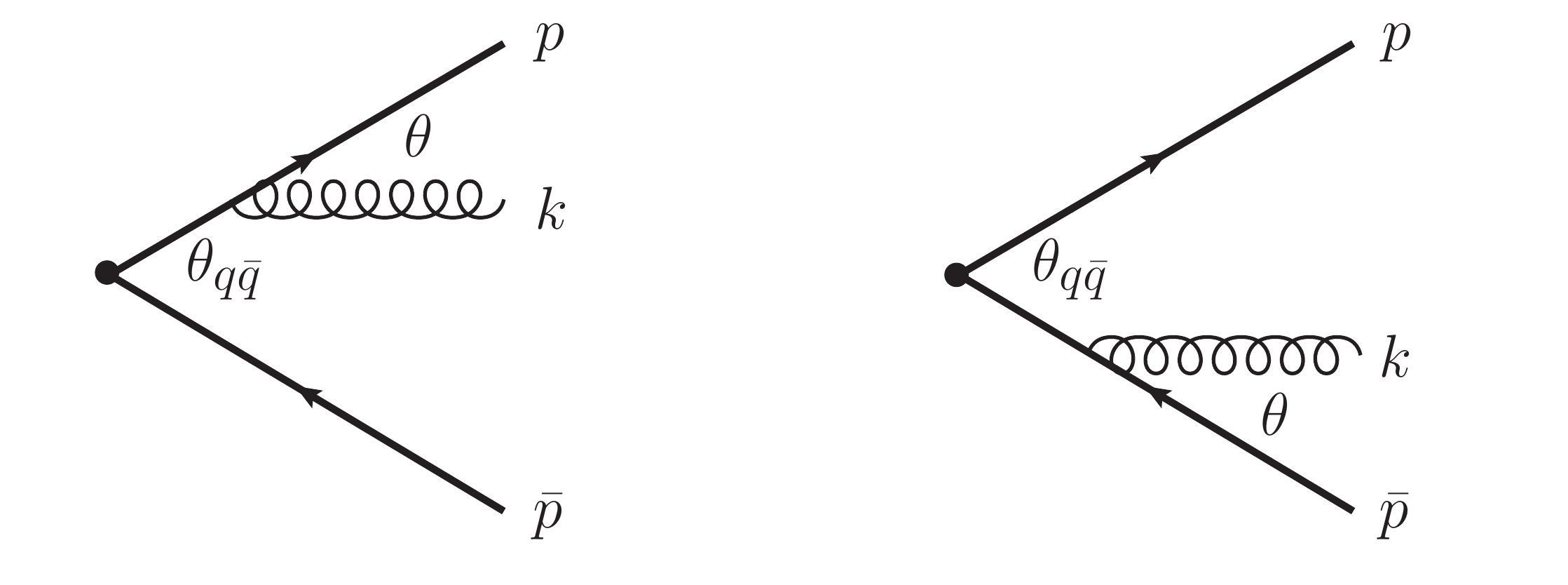}
\caption{Gluon radiation off a $\qqb$-pair. Due to the high virtuality of the initial projectile, i.e., $\gamma^\ast$ or $g^\ast$, we can assume that the pair was created at the origin.}
\end{figure}

The physical setup under consideration is as follows. We assume a virtual, time-like,  photon, $\gamma^\ast$, or gluon, $g^\ast$, that splits into a $\qqb$-pair with momenta $p=(E, {\vec p})$ for the quark and $\bar p = (\bar E, \vec {\bar p})$ for the antiquark, respectively. In the former situation, the antenna starts out in a color singlet state while in the latter, in a color octet one. For highly energetic constituents of the antenna, the splitting occurs on very short time-scales, $t_\text{form} \sim E/Q^2$, where the virtuality of the initial particle is $Q \approx E \theta_\qqb$. Staying within the leading logarithmic approximation, we will therefore only consider gluon emissions which do not resolve the structure of the hard $\qqb$-splitting vertex, i.e., with energies $\omega < E \theta_\qqb$. Nevertheless, for the present purposes, we will always formally consider the limit $E \to \infty$. In the color octet case, for instance, this means that we can altogether neglect radiation that could have occurred before the splitting. In effect, the $\qqb$-pair can be thought of as being created at the origin, $t_0 = 0$, and the emission of the gluon with momentum $k=(\omega, \vec k)$ takes place at much larger times, see fig.~\ref{fig:AntennaVacuum}.

The classical current that describes the $q\bar q$-pair created in vacuum at initial time $t_0=0$ reads 
\be
J^\mu_{(0)}=J^\mu_{q}+J^\mu_{\bar q}+J^\mu_{3},
\ee
where the subscript $(0)$ marks vacuum quantities and the third component, $J_3$, is needed for four-current conservation. Explicitly, the currents for the energetic quark and antiquark, respectively, are given by 
\begin{align}
\labe{eq:currentvac}
J^{\mu,a}_q&= g\frac{p^\mu}{E}~\delta^{(3)}\left(\vec x-\frac{\vec p}{E}t \right)~\Theta(t) ~Q_q^a \,,\\
J^{\mu,a}_{\bar q} &= g\frac{\bar p^\mu}{\bar E}~\delta^{(3)}\left({\vec x}-\frac{\vec{\bar p}}{\bar E}t \right)~\Theta(t)~ Q_{\bar q}^a \,,
\end{align}
where the $Q_q^a$ ($Q_{\bar q}^a$) is the quark (antiquark) color charge, and analogously for $J_3$. In Fourier space the total current reads
\be
J_{(0)}^{\mu,a}(k)=ig\left( \frac{p^\mu}{p\cdot k}Q_q^a\ +\ \frac{\bar p^\mu}{\bar p\cdot k}Q_{\bar q}^a-\frac{p_3^\mu}{p_3\cdot k}Q_3^a\right) \;.
\ee
The continuity relation, $k\cdot J_{(0)}=0$, leads to the conservation of charge, $Q_{q}+Q_{\bar q}=Q_{3}$, and momentum, implying ${\vec p}_3=-{\vec p}-{\vec{\bar p}}$. For a singlet antenna, we simply have $Q_3=0$. In the case of a colored antenna, the third component of the current does not contribute in the frame where $p_3\approx (0,p_3^-,{\bs 0})$ because of the gauge choice. Taking the square of the relation between the charges, namely, $(Q_{q}+Q_{\bar q})^2=0$ and $(Q_{q}+Q_{\bar q})^2=C_A$  for  a singlet and octet antenna respectively, and using the fact that $Q_q^2=Q_{\bar q}^2=C_F$, we easily obtain that 
\be
Q_q\cdot Q_{\bar q} = \left\{ \begin{array}{ll} - C_F & \text{singlet} \\ C_A/2 - C_F & \text{octet} \end{array} \right.
\ee
for our two cases.\footnote{In QCD, the color charges of the quark and gluon are defined as the square of the Casimir operator of the fundamental and adjoint representations, respectively, and read $C_F = (N_c^2-1)/(2N_c)$ and $C_A = N_c$.} These simple relations follow only from conservation of color and hold for arbitrary color representation of the parent. 

Linearizing (\ref{eq:CYMvacuum}) in the coupling, $g$, yields
\be\label{eq:LinVac}
\square A^\mu_{(0)}-\del^\mu (\del\cdot A_{(0)})=J^\mu_{(0)} \;,
\ee
or, more explicitly,
\beq\label{eq:LinVac2}
&&-\del^+ (\del\cdot A_{(0)})=J^+_{(0)}\, ,\nn
&&\square A^-_{(0)}-\del^- (\del\cdot A_{(0)})=J^-_{(0)}\,,\nn
&&\square A^i_{(0)}-\del^i (\del\cdot A_{(0)})=J^i_{(0)}\,.
\eeq
The first equation in (\ref{eq:LinVac2}) is a constraint that relates the various component of the field. Note that only the transverse components of the field are dynamical in LC gauge $A^+=0$, therefore, the second equation can be ignored for gluon production. Hence, plugging the constraint in the last equation in (\ref{eq:LinVac2}), we obtain \citep{MehtarTani:2006xq}
\be
\square A^i_{(0)}=-\frac{\del^i}{\del^+} J^+_{(0)}+J^i_{(0)} \,,
\ee
which, in momentum representation, reads
\beq
\label{eq:field-vac}
-k^2 A_{(0)}^{i,a}(k)=-2\, ig \left( \frac{\kappa^i}{\kh^2}Q_q^a + \frac{\bar\kappa^i}{\khb^2}Q_{\bar q}^a\right) \,.
\eeq
In eq.~(\ref{eq:field-vac}) we have introduced the vectors
\be
\kappa^i=k^i - x\, p^i\,, \qquad  \bar \kappa^i = k^i- \bar x\, \pb^i \,
\ee
which describe the gluon transverse momentum relative to the momentum of the quark or the antiquark, respectively, such that
\be
\kh^2 = 2x \, (p\cdot k) \,, \qquad \khb^2 = 2\bar x \,(\bar p\cdot k) \,.
\ee
Finally, the momentum fractions $x, \bar x$ are defined as $x \equiv k^+/p^+$ and $\bar x \equiv k^+/\bar p^+$. In the limit of small opening angles $x\approx \omega/E$ and $\bar x \approx \omega/\bar E$, which is implicit in the rest of the paper. The soft-gluon emission amplitude is thus connected to the gauge field through the reduction formula (\ref{eq:redform}) via
\be
{\cal M}^a_{\lambda(0)}({\bs k})=\lim_{k^2\to 0} -k^2  A^{i,a}_{(0)}(k) \epsilon_\lambda^i(\bs k) \,.
\ee
The resulting cross section is given by
\beq
d\sigma = d\sigma^{\text{el}}  \sum_{\lambda=\pm 1}  |{\cal M}^a_{(0),\lambda}|^2 \frac{d^3k}{(2\pi)^3\ 2 \omega} \,,
\eeq
and factorizes into the Born elastic cross section for the $\qqb$ production process, $d\sigma^{\text{el}}$, and the radiation described by the square of the emission amplitude. The latter reads
\begin{equation}
\label{eq:M2Vac}
\sum_{\lambda=\pm 1}  |{\cal M}^a_{\lambda(0)}|^2 = 4g^2\left( \frac{1}{\kh^2} Q_q^2 + \frac{1}{\khb^2} Q_{\bar q}^2 + 2\, \frac{\kh\cdot\khb}{\kh^2 \khb^2} Q_q\cdot Q_{\bar q} \right) \,.
\end{equation}
Then, the spectrum of emitted gluons, defined as $dN^\text{vac} = d\sigma/d\sigma^\text{el}$, is readily found in the octet case to be
\beq
\label{eq:spec-general}
\omega\frac{dN^\text{vac}}{d^3k}=\frac{\alpha_s}{(2\pi)^2\,\omega^2}\ \big( C_F\Rcoh+C_A \mathcal{J} \big) \,,
\eeq
where $\Rcoh = \mathcal{R}_q + \mathcal{R}_{\bar q} - 2 \mathcal{J}$. The singlet contribution 
\beq
\mathcal{R}_\text{sing} = 2\omega^2 \frac{p\cdot \bar p}{(p\cdot k)\, (\bar p\cdot k)} \,,
\eeq 
is the well-known antenna emission pattern \citep{Dokshitzer:1991wu}. In eq.~(\ref{eq:spec-general}), we have defined $\mathcal{R}_q \equiv 4 \omega^2/\kh^2$, and analogously for the antiquark, which constitutes the radiation spectrum off an independent constituent, and
\beq
\label{eq:large-angle}
\mathcal{J} \equiv 4\,\omega^2\frac{\kh\cdot \khb}{\kh^2 \khb^2} \,,
\eeq
which describes the quark-antiquark interference. 

To begin with, let us focus on the singlet term $\Rcoh$. This spectrum is divergent when the energy of the emitted gluon becomes soft, $\omega\to 0$. Divergencies also arise when the gluon is emitted collinearly to the either the quark or the antiquark. Introducing the transverse component of the well-known emission current \citep{Dokshitzer:1991wu,bas83}
\beq
\label{eq:emcurr}
\emcurr({\bs k}) \equiv \frac{\kh}{\kh^2} - \frac{ \khb}{\khb^2} \,,
\eeq
the singlet term simply becomes $\Rcoh = 4 \omega^2\, \emcurr^2({\bs k})$. This emission current has a characteristic behavior governed by the characteristic scale related to the opening angle of the pair, namely
\beq
\label{eq:emcurr_vac}
\emcurr^2({\bs k}) = \frac{(k^+ \pip)^2}{\kh^2\, \khb^2}  = \left\{\begin{array}{cc} {\bs k}^{-2} & \qquad \theta \ll \theta_\qqb \,, \\ (k^+ \pip)^2 \,{\bs k}^{-4} & \qquad \theta \gg \theta_\qqb \,, \end{array} \right.
\eeq
where $\theta_\qqb$ is the opening angle of the pair, see fig.~\ref{fig:AntennaVacuum}. This clearly demonstrates that the spectrum is suppressed at large angles due to the presence of destructive interferences.

This point can be further clarified when considering inclusive quantities, i.e., after averaging over the azimuthal angle. It is possible to separate the collinear divergences belonging either to the quark or the antiquark by defining $\Pq \equiv \Rq -\J$ and $\mathcal{P}_{\bar q}\equiv \Rqb-\J$, respectively, so that $\Rcoh = \Pq + \Pqb$. Note that $\Pq$ is divergent along the direction of the quark, $p\cdot k\to 0$, and goes to zero when the gluon is emitted collinearly with respect to the antiquark, $\bar p\cdot k\to 0$. It is therefore interpreted as the probability of a coherent gluon emission off the quark, and implies strict angular ordering. Namely, by setting the quark momentum on the $z$ axis and integrate over the azimuthal angle we obtain 
\be
\int_0^{2\pi} \frac{d\varphi}{2\pi}\ \Pq= \frac{2}{1-\cos\theta}~\Theta(\cos\theta-\cos\theta_\qqb)\,,
\ee
where $\theta$ is the angle between the quark and the emitted gluon, see fig.~\ref{fig:AntennaVacuum}. Hence, after averaging over the azimuthal angle, gluon emissions off the quark are confined within the cone with opening angle $\theta_\qqb$ and centered on the quark direction. The corresponding emission spectrum off the quark reads  
\be
\label{eq:spectrum-vac2}
dN^\text{vac}_q=\frac{\alpha_s C_F}{\pi} \frac{d\omega}{\omega}\frac{\sin\theta d \theta}{1-\cos\theta} \ \Theta(\cos\theta-\cos\theta_\qqb) \,.
\ee
In addition to the double, soft ($\omega\to 0$) and collinear ($\theta\to 0$), logarithmic singularity the coherent spectrum contains the key feature of angular ordering. The coherent spectrum off the antiquark is found analogously. This procedure serves to establish a probabilistic picture of jet fragmentation for the purpose of describing inclusive jet characteristics.

In the octet case, the latter term in eq.~(\ref{eq:spec-general}), coming with the total color charge of the pair, $C_A$, gives rise to large-angle radiation. In fact, averaging it over the azimuthal angle along the direction of the quark we get
\beq
\label{eq:antiangularordering}
\int_0^{2\pi} \frac{d\varphi}{2\pi}\ \J= \frac{2}{1-\cos\theta}~\Theta(\cos\theta_\qqb-\cos\theta) \,. 
\eeq
Since the large-angle radiation appears with the total color charge of the system, it is therefore re-interpreted as radiation off the initial gluon {\it imagined} to be on shell \citep{Dokshitzer:1991wu}. Thus, in vacuum, large-angle gluon emission is sensitive to the total charge of the system. This property relies on color conservation and leads to angular ordering of the vacuum cascade in QCD.

\section{Antenna radiation in medium -- the color singlet case}
\label{sec:medium}
\begin{figure}
\centering
\includegraphics[width=0.7\textwidth]{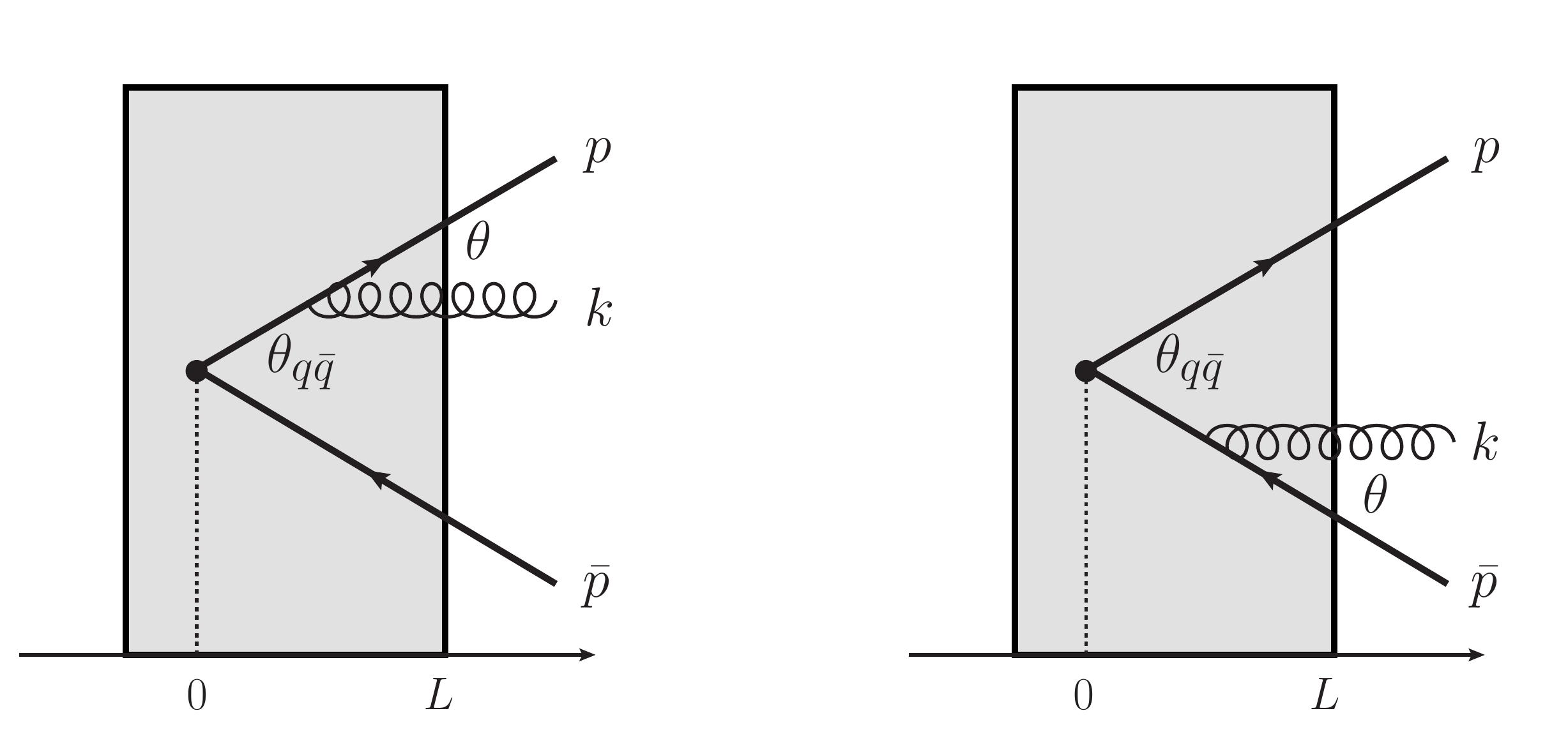}
\caption{Diagrammatic illustration of in-medium gluon radiation off a $\qqb$ pair.}
\label{fig:AntennaMedium}
\end{figure}
Let us turn now to calculate the medium modification of the gluon spectrum off the $\qqb$ antenna.  For the sake of clarity, we assume for the moment that the pair is initially in a color singlet state, i.e., $Q_q + Q_{\bar q} = 0$, leaving us only with a medium modification of the first term in eq.~(\ref{eq:spec-general}). A generalization to the color octet initial configuration will be done in Section~\ref{sec:octet}.

We assume that the pair is collimated in the $+z$ direction while the medium propagates in the opposite, $-z$, direction at nearly the speed of light. At the end of the calculation we boost back to the lab rest frame, where the medium is at rest. 
Therefore, this approximation is only valid as long as the pair opening angle $\theta_\qqb\ll 1$ and $E\to \infty$. This allows us to pick up the high energy limit. The process under consideration in sketched in fig.~\ref{fig:AntennaMedium}. We shall treat the pair field as a perturbation around the strong medium field, denoted by $A_\sM$, thus, the total field can express as 
\be
A^\mu\equiv A^\mu_\sM+A^\mu_{(0)}+A^\mu_{(1)},
\ee
where the $A_{(0)}$, calculated in the previous section, is the gauge field of the pair in the absence of the medium, and $A_{(1)}$ is the linear response of the medium to the perturbation caused by the pair. The subscript $(1)$ marks the response of the field at first order in $A^-_\text{med}$. In the asymptotic limit, the medium gauge field is solution of the 2-dimensional Poisson equation
\be
-\bdel^2A^-_\sM(x^+,\x)=\rho_\sM(x^+,\x)\,,\qquad A_\sM^i = A^+_\sM = 0 \,,
\ee
where $\rho_\sM(x^+,\x)$ describes the static distribution of medium color charges. 
Thus, the only non-zero component reads, in Fourier space,
\be
A^-_\sM (q) = 2\pi\ \delta(q^+)\ \int \!\! dx^+\ \A_\sM(x^+,\q)~e^{iq^-x^+} \,.
\ee
Thus,  at leading order in the medium field, the CYM equations read
\beq
\label{eq:field-1}-\del^+ \left( \del\cdot A_{(1)} \right) &=& J^+_{(1)}\,,\\
\label{eq:field-2}\square A_{(1)}^--2ig \left[ A^-_\sM, \, \del^+ A_{(0)}^- \right] &=& \del^- \left( \del\cdot A_{(1)} \right)+J_{(1)}^-Ê\;,\\
\label{eq:field-3}\square A_{(1)}^i-2ig \left[ A^-_\sM, \, \del^+ A_{(0)}^i \right] &=&\del^i \left( \del\cdot A_{(1)} \right)+J_{(1)}^iÊ\;.
\eeq
Again, as for the vacuum case we can drop the redundant equation for the negative light-cone component of the field, then using the constraint (\ref{eq:field-1}) in eq. (\ref{eq:field-3}), we obtain \citep{MehtarTani:2006xq,Gelis:2005pt}
\beq
\label{eq:CYMmedium}
\square A_{(1)}^i-2ig \left[ A^-_\sM,\del^+ A_{(0)}^i \right] = -\frac{\del^i}{\del^+}J_{(1)}^++J_{(1)}^iÊ\,.
\eeq
Furthermore, the current obeys the continuity relation, 
\be
\del_\mu J_{(1)}^\mu=ig \left[ A^-_\sM,J_{(0)}^+ \right]  \;,
\ee
which can be solved by 
\be
J^\mu_{(1)}=ig\frac{p^\mu}{p\cdot \del }~ \left[ A^-_\sM,J^+_q \right] + ig \frac{ \pb^\mu}{ \pb\cdot \del }~ \left[ A^-_\sM,J^+_{\bar q} \right] \;.
\ee
We shall now focus on the quark contribution to the gluon emission amplitude. The anti-quark part is calculated analogously and added at the level of the amplitude. In Fourier space, we have
\be
J_{q(1)}^{\mu,a}(k)=(ig)^2\frac{p^\mu}{-i\,p\cdot k }\int\!\! \frac{d^4q}{(2\pi)^4}~\frac{p^+}{p\cdot (k-q)+i\epsilon}~i \left[ T\cdot A^{-}_\sM(q)\right]^{ab}Q_q^b \,,
\ee
where  $[T\cdot A_\text{med}(q) ]^{ab}Q^b_q = -i f^{abc} A^c_\text{med}(q) Q^b_q$ and $f^{abc}$ is the SU(3) structure constant. The $q^+$ integral is trivial thanks to $\delta(q^+)$ in $A^-_\sM$. The $q^-$ integral is equal to the residue at the pole 
\be
q^-=k^-+\frac{p^-k^+-\pp\cdot \Q}{p^+}+i\epsilon \,, 
\ee
yielding 
\begin{multline}
J_{q(1)}^{\mu,a}(k) = -ig^2\frac{p^\mu}{p\cdot k} \int\!\! \frac{d^2\q}{(2\pi)^2} \int_{0}^{\infty}\!\! dx^+ e^{i \frac{p^+k^- + p^- k^+ - \pp\cdot\Q}{p^+} x^+}  \left[ T\cdot \A_\sM(x^+,\q) \right]^{ab} \!Q_q^b \,.
\end{multline}
Inserting this expression into (\ref{eq:CYMmedium}) gives
\begin{multline}
\label{eq:field1}
-k^2A_{q(1)}^i(k) = 2g \int \frac{d^4q}{(2\pi)^4}\, \left[A_\sM^-(q), (k-q)^+ A^i_{q(0)} (k-q) \right] \\  -\frac{k^i}{k^+}J^+_{q(1)}(k)+J^i_{q(1)}(k) \,,
\end{multline}
in momentum space. From eq.~(\ref{eq:field-vac}) we read off the quark-induced vacuum field, namely
\be
-k^2 A^{i,a}_{q(0)}(k)\;=\;-2\, ig\frac{\kappa^i}{\kh^2}\,Q_q^a \;.
\ee
The $q^+$-integral in eq.~(\ref{eq:field1}) is again trivial while the $q^-$-integration picks up the contribution from two poles. In the retarded prescription the first pole appears at $  (k-q)^2+ik^+\epsilon =0$, such that
\be
q^-=k^- - \frac{\Q^2}{2k^+} + i\epsilon \;,
\ee
and the second at $p\cdot  (k-q) + i\epsilon=0$, where
\be
q^-=k^- + \frac{p^-k^+-\pp\cdot\Q}{2k^+}+i\epsilon \;. 
\ee
After performing the integrations we find the induced field from the quark part to be given by
\begin{multline}
-k^2 A^{i,a}_{q(1)}(k) = 2\,ig^2 \int \frac{d^2 \q}{(2\pi)^2} \int_{0}^{\infty}\!\! dx^+ \left[ T\cdot\A_\sM (x^+,\q)\right]^{ab}Q^b_q ~e^{i\left( k^- - \frac{({\bs k} - \q)^2}{2 k^+} \right)x^+} \\
\times\left\{ \frac{(\kappa- q)^i}{(\kh-\q)^2} \left[1 - \exp \left( i \frac{(\kh -\q)^2}{2k^+} x^+ \right) \right] + \frac{\kappa^i}{\kh^2} \exp \left( i \frac{(\kh - \q)^2}{2k^+} x^+ \right) \right\}  .
\end{multline}
Thus, the amplitude for gluon radiation off the quark reads
\begin{multline}
\label{eq:amplitude}
{\cal M}^{a}_{\lambda, q(1)} = 2\,ig^2 \int \frac{d^2 \q}{(2\pi)^2} \int_{0}^{\infty}\!\! dx^+ \left[ T\cdot\A_\sM(x^+,\q)\right]^{ab}Q^b_q \\
 \times\left\{ \frac{\kh - \q}{(\kh - \q)^2} - {\bs L} \,\exp \left[ i \frac{(\Qh)^2}{2k^+} x^+ \right] \right\} \cdot \epsp\;,
\end{multline}
where 
\be\label{eq:lipatov}
 {\bs L}= \frac{\kh - \q}{(\Qh)^2} -\frac{\kh}{\kh^2} 
\ee
is the transverse component of the well-known Lipatov vertex in LC gauge \citep{Baier:1996kr,Kuraev:1977fs,Balitsky:1978ic}. The amplitude for gluon radiation off the anti-quark, $\mathcal{M}_{\bar q(1)}$, is deduced from $\mathcal{M}_{q(1)}$ by substituting the quark momentum $p\to \bar p$ and charge $q \to \bar q$. Note that we have dropped an overall phase factor in eq.~(\ref{eq:amplitude}) which cancels in the cross section.

The Lipatov term in eq.~(\ref{eq:amplitude}) corresponds to the induced radiation off an asymptotical (on-shell) particle. The other term represents the bremsstrahlung off an accelerated particle with a subsequent rescattering of the real, emitted gluon. Furthermore, note that when the gluon becomes collinear to the quark before rescattering, i.e., $(\Qh)^2 \to 0$, no singularity arises because of the phase structure. 

Let us now turn to the evaluation of the cross-section. The gluon spectrum is calculated by taking the square of the amplitude and averaging over the medium field. To do so, we assume the medium color charges to be uncorrelated along the $x^+$-direction and having longitudinal support on the line element $[0,L^+]$ but infinite and uniform in the transverse direction. Thus, one can treat the medium charge density as a Gaussian white noise defined by the two-point function 
\be
\langle \rho^a_\sM(x^+,\q) \rho^{\ast b}_\sM(x'^+,\q')\rangle = \delta^{ab} m_D^2~ n(x^+)\delta(x^+-x'^+)~(2\pi)^2 \delta^{(2)}(\q-\q') \;,
\ee
which yields 
\begin{multline}
\label{eq:MediumAverage}
\langle \A^a_\sM(x^+,\q) \A^{\ast b}_\sM(x'^+,\q')\rangle = \delta^{ab} m_D^2 n(x^+) \delta(x^+-x'^+)(2\pi)^2 \delta^{(2)}(\q-\q') \, {\cal V}^ 2(\q) \,,
\end{multline}
for the medium average of the interaction potential. Following previous works, we define the potential $\mathcal{V}(\q)$ to be of the Yukawa-type,
\be
\label{eq:MediumPotential}
{\cal V}(\q) \;=\; \frac{1}{\q^2+m_D^2} \;,
\ee
with $m_D$ being an infrared cut-off usually identified with the in-medium Debye screening mass. To simplify the discussion in what follows, we assume the medium to be uniform in the longitudinal direction, such that the one-dimensional medium density is constant, i.e., $n(x^+)= n_0 \Theta(L^+-x^+)$, where $L = L^+/\sqrt{2}$ is the medium size in the longitudinal direction. Hereafter, we will also define the medium transport parameter $\hat q$ as 
\beq
\label{eq:qhat}
\hat q = \alpha_s C_A \,n_0 m_D^2 \;.
\eeq
Note that this definition differs slightly from the one used in ref.~\citep{sal03}.\footnote{The standard definition of $\hat q$ refers to a different approximation scheme, the multiple soft scattering approximation. Here, it is meant as a shorthand.}

After averaging over medium configurations and summing over gluon polarizations the spectrum finally reads
\beq
\label{eq:SpectrumMed}
\omega\frac{dN^\text{med}}{d^3k}&=&\frac{8Ê\,\alpha_s C_F \,\hat q}{\pi} \potint ~\int_0^{L^+} dx^+\nn
&&\left\{\left[1-\cos\left(\frac{\kh+\khb - 2\q}{2}\cdot\pip\, x^+\right)\right] {\bs L}\cdot \bar{\bs L}\right.\nn
&&+ \left[1-\cos\left(\frac{(\Qh)^2}{2k^+} x^+\right)\right] \emcurr({\bs k}-\q) \cdot {\bs L}\nn
&&\left.- \left[1-\cos\left(\frac{(\Qhb)^2}{2k^+} x^+\right)\right] \emcurr({\bs k} -\q) \cdot \bar {\bs L}\right\} \,,
\eeq
 where we have introduced the integration measure
\beq
\potint \equiv \int \frac{d^2 \q}{(2\pi)^2} \mathcal{V}^2(\q) \,,
\eeq
to ease the notation. The spectrum depends on two emission currents: 1) the Lipatov vertices ${\bs L}$ and $\bar {\bs L}$, given in eq.~(\ref{eq:lipatov}), and 2) the transverse emission current coming from the hard emission vertex, first term in eq.~(\ref{eq:amplitude}), associated with gluon rescattering, which reads
\beq
\label{eq:emcurr2}
\emcurr({\bs k}-\q) \equiv \frac{\kh -\q}{(\Qh)^2} - \frac{\khb - \q}{(\Qhb)^2} \,.
\eeq
This current has the same structure as the corresponding vacuum current $\emcurr({\bs k})$, see eq.~(\ref{eq:emcurr}), apart from the characteristic shift of the momentum ${\bs k} \to {\bs k} - \q$ due to the medium rescattering.

The products of emission currents in eq.~(\ref{eq:SpectrumMed}) come with the corresponding formation time phase structures. The two latter lines in eq.~(\ref{eq:SpectrumMed}) depend on the formation time of the rescattered gluon emitted off either of the antenna constituents which, as will be shown in Section~\ref{sec:glv}, is identical to the phase structure of their respective independent emission spectra. The phase related with first term, proportional to ${\bs L}\cdot \bar {\bs L}$, is in fact the difference between the former two, since 
\beq
\frac{(\Qh)^2}{2k^+} - \frac{(\Qhb)^2}{2k^+} = \frac{\kh+\khb - 2 \q}{2}\cdot\pip \,,
\eeq
where
\be
\label{eq:deltan}
\pip\equiv \frac{\kh-\khb}{k^+}\,.
\ee
Note that the vector $\pip$ is closely related to the opening angle of the pair, since  $\left| \pip \right| \sim \sin\theta_\qqb$.
 
To clarify further how to obtain (\ref{eq:SpectrumMed}) from the amplitude (\ref{eq:amplitude}), we emphasize that virtual corrections, often called contact terms \citep{Baier:1996kr,Baier:1996sk,Wiedemann:2000za}, have been added to the square of the amplitude (\ref{eq:amplitude}). These terms account for unitarity and read 
\beq
\label{eq:ContactTermDefinition}
\omega \left.\frac{dN^{\text{med}}}{d^3k}\right|_\text{virtual} = -\frac{2\,\alpha_s C_F}{\pi}  \frac{p\cdot \pb}{(p\cdot k)(\pb\cdot k)} \, \hat q L^+\int \frac{d^2 \q}{(2\pi)^2}  {\cal V}^2(\q) \,.
\eeq
In other words, the inclusion of the contact terms simply corresponds to redefining the potential as
\be
\label{eq:ContactTermDef}
{\cal V}^2(\q)\to{\cal V}^2(\q)-(2\pi)^2\delta^{(2)}(\q)\int\frac{d^2 \q'}{(2\pi)^2}{\cal V}^2(\q') \;,
\ee
which insures that (\ref{eq:SpectrumMed}) is finite in the $\q\to0$ limit.

To make contact between the semi-classical calculations above with more common approaches to QCD, we have also calculated the corresponding set of amplitudes for the process under consideration using Feynman rules, c.f. eq.~(\ref{eq:amplitude}), in Appendix~\ref{sec:feynman} and cross sections in Appendix~\ref{sec:feynman-cs}.

\section{The gluon independent spectrum off the quark}
\label{sec:glv}
Before discussing the full structure of eq.~(\ref{eq:SpectrumMed}) in detail, we focus initially only on the terms where the gluon is emitted and absorbed by the same quark in amplitude and complex-conjugate amplitude, respectively. In other words, the spectrum is proportional to $\left| \mathcal{M}_{q(1)} \right|^2$ for the quark and, therefore, does not contain any information about the antiquark, and vice versa. The independent spectrum off the quark reads
\beq
\label{eq:GLV1}
\omega \frac{dN^\text{indep}_q}{d^3 k} =  \frac{8\, \alpha_s C_F\, \hat q}{\pi} \potint \int_0^{L^+}\!\! dx^+\left[ 1-\cos\frac{(\Qh)^2}{2k^+} x^+ \right] \frac{\kh \cdot \q}{(\Qh)^2 \kh^2} \,,
\eeq
and similarly for the antiquark. In the special case when the quark is traveling along the $z$-direction, i.e., $|{\bs p}|=0$ and $\kh = {\bs k}$, the spectrum (\ref{eq:GLV1}) becomes the familiar GLV \citep{gyu00} or BDMPS-Z  spectrum at first order in medium opacity \citep{Wiedemann:2000ez,Wiedemann:2000za} off a fast moving quark. The spectrum in eq.~(\ref{eq:GLV1}) is infrared and collinear finite \citep{sal03}. The generalization to multiple scattering with the medium, also extensively used in phenomenological investigations, is known as well \citep{Baier:1996kr,Baier:1996sk,Zakharov:1996fv,Zakharov:1997uu,Wiedemann:2000ez,Wiedemann:2000za}. 

Let us at present briefly recall some key features of this spectrum which can aid our understanding of the antenna dynamics. For the time being, we will work in the $|{\bs p}| = 0$ frame. The argument of the cosine in eq.~(\ref{eq:GLV1}) relates the effective gluon formation time
\be
t_\text{form} \sim \frac{2\,\omega}{({\bs k} - \q)^2}
\ee 
to the position of the interaction with the medium. Comparing $t_\text{form}$ to the length of the medium reveals an interplay between two emission mechanisms that are related to the diagrammatic interpretation described in the previous subsection. Firstly, we notive that the spectrum achieves the maximal emission rate when $t_\text{form} \ll L$. For typical momenta of the order of the medium scale, $({\bs k} - \q)^2 \sim m_D^2$, the previous condition translates to $\omega \ll \bar \omega_c$, where
\beq
\label{eq:omegac}
\bar \omega_c \equiv \frac{1}{2}m_D^2 L \,.
\eeq
In this case, the cosine in eq.~(\ref{eq:GLV1}) oscillates and the term proportional to it is suppressed. The remainder can be written as
\beq
\label{eq:GLVprobabilistic}
\omega \left. \frac{dN^\text{indep}_q}{d^3 k}\right|_{\omega \ll \bar \omega_c} = \frac{4~ \alpha_s C_F\, \hat q L^+}{\pi}  \potint \left[ {\bs L}^2 + \frac{1}{({\bs k}-\q)^2} -\frac{1}{{\bs k}^2} \right] \,,
\eeq
which permits a clear probabilistic interpretation \citep{Wiedemann:2000za}. The latter two terms correspond to the bremsstrahlung of the accelerated charge which might undergo further rescatterings. While the last factor in eq.~(\ref{eq:GLVprobabilistic}) serves to reduce the rate of collinear emissions, $\sim 1/{\bs k}^2$, in vacuum, the second factor is associated with the radiation component which rescatters once in the medium. This is indeed only a reshuffling of the transverse momentum of the real emitted gluons, ${\bs k} \to {\bs k} -\q$, and do not affect the total multiplicity provided $m_D \ll \omega$. 

The first term in eq.~(\ref{eq:GLVprobabilistic}) represents the genuine induced emission due to the hard scattering of an asymptotic charge with the medium. Explicitly, it reads
\beq
\label{eq:GunionBertsch}
{\bs L}^2 = \frac{\q^2}{{\bs k}^2 ({\bs k}-\q)^2} \,,
\eeq
which is just the Gunion-Bertsch spectrum \citep{Gunion:1981qs}. This decomposition can be extended to arbitrary order in medium opacity \citep{Wiedemann:2000za}. Later, we will come back to how this interpretation is generalized in the antenna case.

For larger formation times, $t_\text{form} \gtrsim L$, the phase structure spoils the separation of the two emission mechanisms described above and leads to a destructive interference between them. This is analogous to the so-called Landau-Migdal-Pomeranchuk (LPM) effect which describes the coherent interaction with the medium scattering centers and leads to the suppression of the spectrum, see below.

\begin{figure}
\centering
\includegraphics[width=0.8\textwidth]{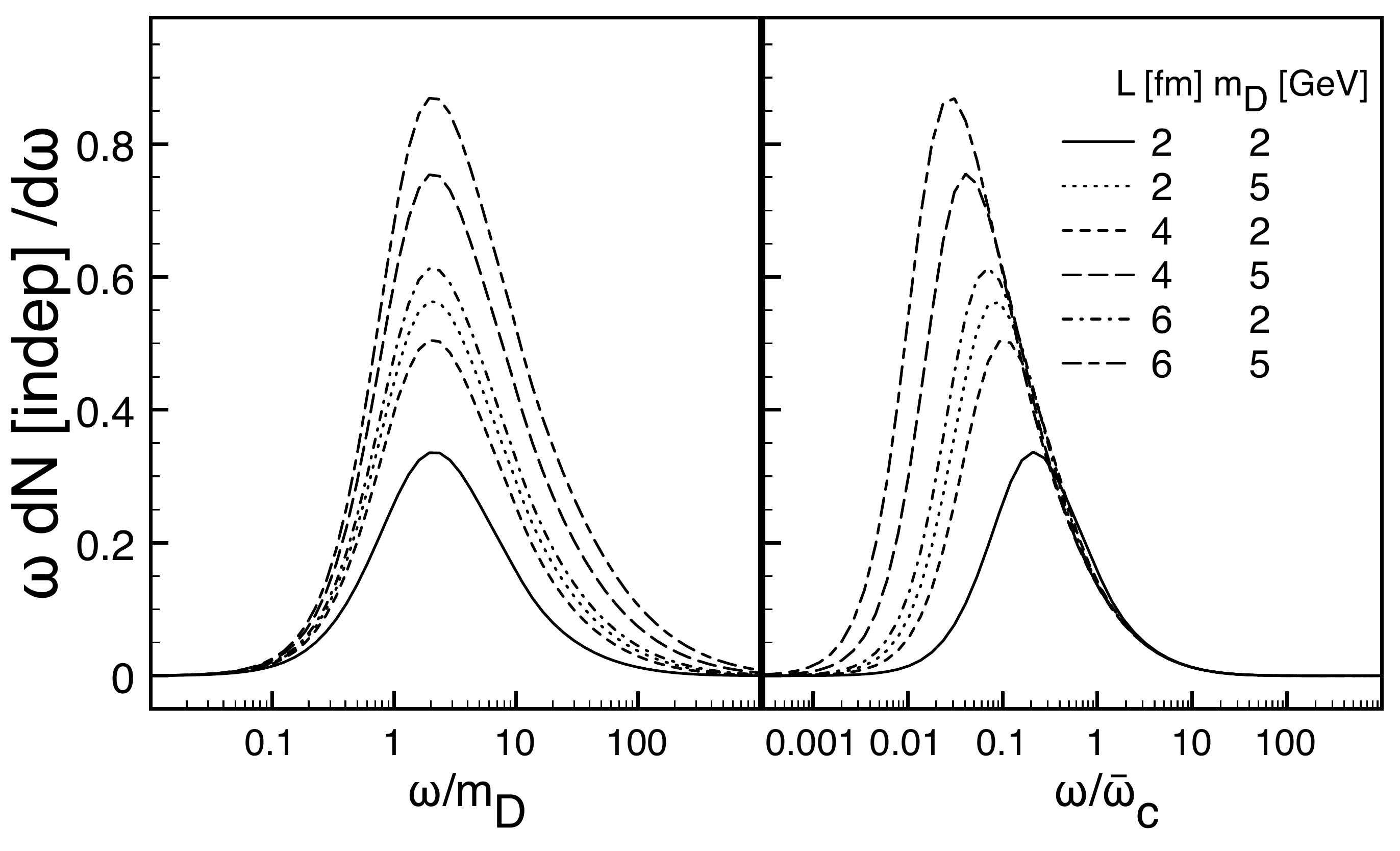}
\caption{The medium-induced independent spectrum off a quark, given by eq.~(\ref{eq:GLV1}) integrated over angles, for a set of medium parameters: $L \text{ [fm]} = \{2, 4, 6 \}$ and $m_D \text{ [GeV]} = \{ 2, 5\}$.}
\label{fig:GLV}
\end{figure}
Integrating out the transverse momenta in eq.~(\ref{eq:GLV1}), the independent spectrum, written as $\omega dN^\text{indep}_q/d\omega$, has the following general features \citep{sal03}
\beq
\label{eq:GLVdetails}
\omega \frac{dN^\text{indep}_q}{d\omega} \sim \frac{\hat q L^+}{m_D^2} \left\{ \begin{array}{cc}  \log \frac{\bar \omega_c}{\omega} & \quad m_D < \omega < \bar \omega_c \,, \\ \frac{\bar \omega_c}{\omega} & \quad \bar \omega_c < \omega \,. \end{array} \right.
\eeq
For smaller energies, $\omega < m_D$, the spectrum is strongly suppressed, revealing the infrared safety. Because of this change of behavior around the hard scale of the medium, a maximum of the spectrum is expected at $m_D$. For $\omega > m_D$ the spectrum is continually dropping.

These features are evident in the numerical evaluation of (\ref{eq:GLV1}), depicted in fig.~\ref{fig:GLV} for a set of medium parameters, $m_D$ and $L$ (the values are given in the figure caption). Firstly, in the left panel we see the peak behavior follows the typical transverse scale of the medium. The  scaling with the characteristic gluon energy $\bar \omega_c$ controls the tail of the distribution, see the right panel in fig.~\ref{fig:GLV}.

Keeping in line with the discussion of the vacuum radiation in Section \ref{sec:vacuum}, the presently discussed component of the total spectrum is the medium-induced analog of the independent spectrum off the quark in the vacuum, and we thus denote it
\beq
\label{eq:SpecGLV}
\omega \frac{dN^\text{indep}_q}{d^3 k} = \frac{\alpha_s C_F}{(2\pi)^2 \, \omega^2} \mathcal{R}^\text{med}_q \;,
\eeq
and similarly for the antiquark. We come back to how it compares to the full coherent spectrum in the presence of another emitter in Section \ref{sec:numerics}.

\section{The soft limit}
\label{sec:soft}
In addition to the diagrams described in the previous subsection, where the gluon is emitted and subsequently absorbed by the same antenna constituent, we also find novel contributions stemming from a medium-induced interference between the two emitters of the antenna. These contributions were first discussed in ref.~\citep{MehtarTani:2010ma}. Therefore, let us here recall some general features of the medium-induced spectrum (\ref{eq:SpectrumMed}) in the soft limit, i.e., for $\omega \to 0$, where it simplifies considerably allowing for a completely analytical treatment. The leading contribution arises from the first term in eq.~(\ref{eq:SpectrumMed}), where
\beq
\label{eq:lipatovsoft}
\lim_{\omega\to 0} {\bs L}\cdot \bar{\bs L}=\frac{\kh\cdot\khb}{\kh^2\khb^2} \,.
\eeq
Thus, it is immediately evident that the emissions will be vacuum-like since this simply equals $\mathcal{J} / (4 \omega^2)$ according to the definition in eq.~(\ref{eq:large-angle}). As previously discussed, this term is suppressed inside the cone and becomes zero after azimuthal integration, cf. eq.~(\ref{eq:antiangularordering}). Furthermore, the argument of the cosine simplifies to $\lim_{\omega \to 0}(\kh+\khb - 2 \q)/2 = -\q$. The spectrum reads then 
\beq
\label{eq:SpectrumMedSoft}
\omega\frac{dN^{\text{med}}}{d^3k} = \frac{\alpha_s C_F}{\pi\,\omega^2}\ 
2\mathcal{J}
\int_0^{L^+}\!\! d x^+  ~\hat q \,\sigma \left(|\pip|x^+ \right) \,.
\eeq
The forward dipole--medium amplitude, appearing in eq.~(\ref{eq:SpectrumMedSoft}), is given by 
\begin{align}
\label{eq:DipoleCrossSection}
\sigma(|\delta{\bs n}| x^+) &= \potint \left(1-\cos \pip\cdot \q  \,x^+ \right)\nn
&= \frac{1}{4\pi m_D^2}\bigg[1 - \frac{r_\perp \, m_D \, x^+}{L^+} K_1 \bigg(\frac{r_\perp \, m_D \, x^+}{L^+}\bigg)\bigg] \,, 
\end{align}
where 
\beq
\label{eq:rperp}
\r \equiv \pip \, L^+ \,,
\eeq
and $r_\perp \equiv |\r| \simeq \theta_\qqb L$ is the antenna-dipole size probed by the total medium length.\footnote{We adopt the similar notation for the absolute value of perpendicular vectors in what follows, i.e., $x_\perp \equiv |{\bs x} |$.}

Most importantly, eq.~(\ref{eq:SpectrumMedSoft}) contains a soft divergency and vanishes if the gluon is collinear to either the quark or the antiquark. A further remarkable property of eq.~(\ref{eq:SpectrumMedSoft}) is the factorisation of the radiation process, described entirely by $2\mathcal{J}$, and the medium interaction, which is fully contained in the product $\hat q\, \sigma(|\delta{\bs n}| x^+)$.

Writing the phase space out in detail, we obtain the total medium-induced antenna spectrum in the soft limit
\beq
\left. dN^\text{med}\right|_{\omega\to 0} = \frac{\alpha_s C_F}{\pi} \Delta_\text{med}(\theta_\qqb,L) ~2\J ~\frac{d\omega}{\omega} \frac{d \Omega}{4\pi} \,,
\eeq
where
\be
\label{eq:amed}
\Delta_\text{med}(\theta_\qqb,L) \equiv \frac{{\hat q}}{m_D^2} \int_0^{L^+} {\rm d} x^+ \, \bigg[1 - \frac{r_\perp m_D \, x^+}{L^+} K_1 \bigg(\frac{r_\perp m_D \, x^+}{L^+}\bigg)\bigg] \,,
\ee
and $d\Omega = d \cos \theta \,d\varphi$. The quantity in eq.~(\ref{eq:amed}) can be interpreted as a decoherence parameter, as we shall discuss in detail below. Proceeding as for the calculation of the spectrum in vacuum, see Section~\ref{sec:vacuum}, the medium-induced angular radiation pattern off the quark is simply given by
\be
\lim_{\omega\to 0}{\cal P}^\text{med}_q = \lim_{\omega \to 0} \left(\mathcal{R}_q^\text{med} - \mathcal{J}^\text{med}_q \right) =  \J \,,
\ee
and analogously for the antiquark, since the medium-induced independent spectra are vanishing in the soft limit. Thus, after integrating out the azimuthal angle $\varphi$, we obtain
\beq
\label{eq:nqmed}
\left. dN^{\text{med}}_q\right|_{\omega\to 0}=\frac{\alpha_sC_F }{\pi}\,\Delta_{\text{med}}(\theta_\qqb,L)~\frac{d\omega}{\omega}\frac{\sin\theta \ d \theta}{1-\cos\theta} \Theta(\cos\theta_{q\bar q}-\cos\theta) \;.
\eeq
demonstrating that medium-induced soft gluon radiation is suppressed inside the cone of opening angle $\theta_{q\bar q}$, as opposed to the standard angular structure obtained in vacuum, see eq.~(\ref{eq:spectrum-vac2}). The medium parameters only enter in the decoherence parameter $\Delta_\text{med}$, which does not depend on $\theta$, such that the functional form of the spectrum remains vacuum-like and antiangular ordered \citep{MehtarTani:2010ma}, see eq.~(\ref{eq:antiangularordering}). 

The spectrum found above has some similarities with the radiation off a color octet antenna in the vacuum, c.f. eqs.~(\ref{eq:spec-general}) and (\ref{eq:antiangularordering}), in which a large-angle contribution, corresponding to radiation off the total charge of the pair, also appears. Several crucial differences exist, however. Most importantly, (i) the disappearance of the medium-induced radiation in the limit of vanishing opening angles, $\theta_\qqb \to 0$, as expected of a singlet antenna and (ii) the relevant color factor, $C_F$, in the medium case, indicating the radiation off a quark in the fundamental representation. The latter point is, in fact, further clarified by taking into account multiple  scattering with the medium. Then, the adjoint color $C_A$ factor, contained in $\hat q$, exponentiates \citep{MehtarTani:2011tz}.

\begin{figure}
\centering
\includegraphics[width=0.65\textwidth]{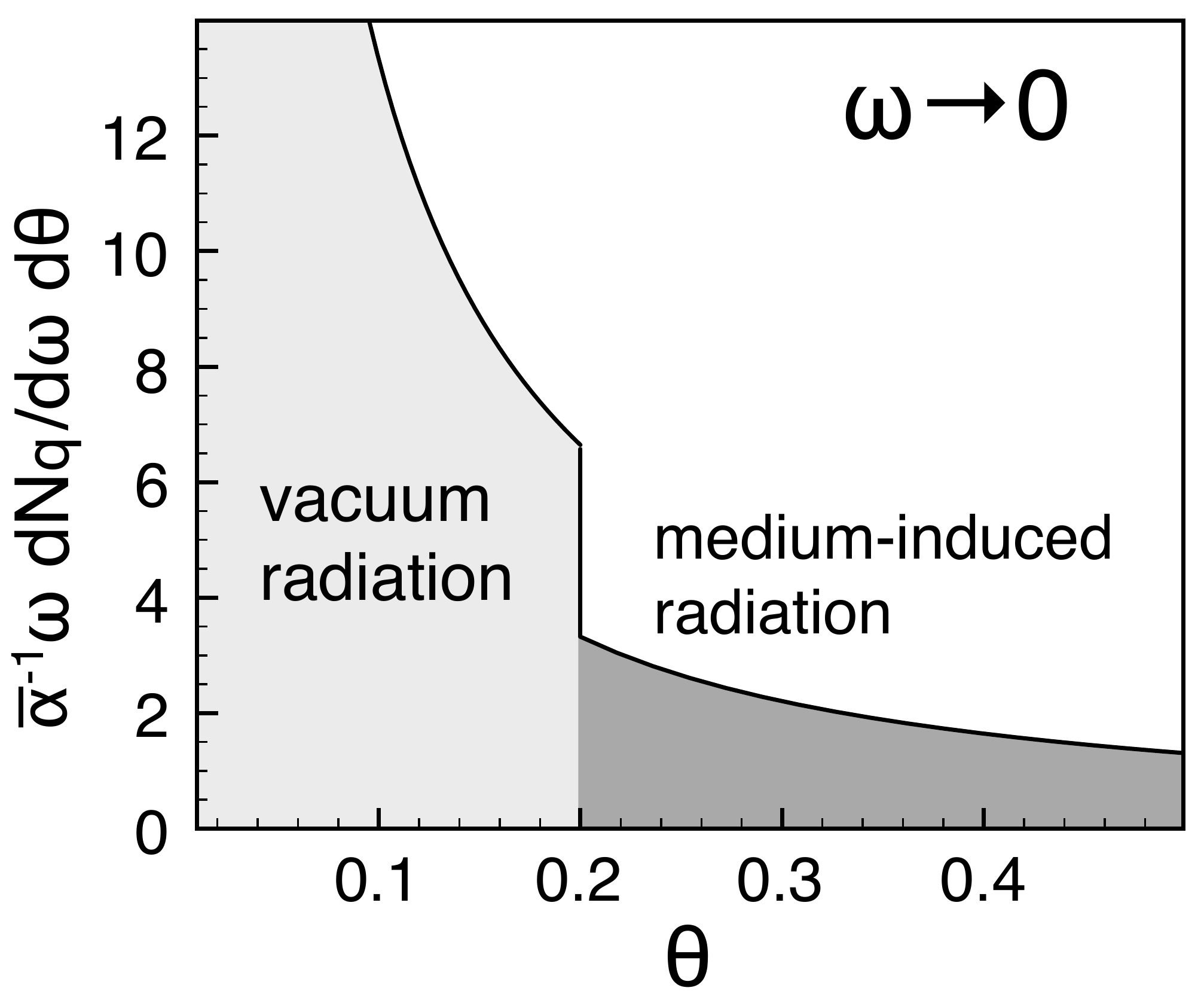}
\caption{The inclusive soft gluon emission spectrum off a $\qqb$-antenna constituent in the presence of a QCD medium, according to (\ref{eq:nqtotal}) after azimuthal average. The parameters are chosen to be $\theta_\qqb = 0.2$ and $\Delta_\text{med} = 0.5$.}
\label{fig:VacMedIllustration}
\end{figure}
Thus, the behavior of eq.~(\ref{eq:nqmed}) reflects the partial decoherence of the quark and the antiquark radiation due to their in-medium color rotations controlled by the decoherence parameter $\Delta_\text{med}$. In summary, the gluon spectrum off a $\qqb$-antenna constituent in the presence of a medium  in the soft limit reads
\be
\label{eq:nqtotal}
\left. dN_q^{\text{tot}} \right|_{\omega \to 0}= \frac{\alpha_sC_F }{\pi}~\frac{d\omega}{\omega}\frac{\sin\theta \ d \theta}{1-\cos\theta}\Big[ \Theta(\cos\theta-\cos\theta_\qqb)+ \Delta_{\text{med}}\Theta(\cos\theta_{q\bar q}-\cos\theta)\Big] \,.
\ee
For a general expression, also valid for an antenna in a color octet state, see eq.~(\ref{eq:spec-medium-octet}). The medium-induced component leads to a gradual onset of large-angle gluon emissions with increasing medium density and/or length. These simple features are illustrated in fig.~\ref{fig:VacMedIllustration}. 

In the following we establish how this interpretation generalizes for finite gluon energies. At the outset, we note that two distinct regimes can be identified from the functional analysis of the decoherence parameter appearing in eq.~(\ref{eq:nqmed}), which will prove essential in the discussion of characteristics of the full spectrum.
\begin{itemize}
\item {\it The ``dipole" regime:} where $r_\perp \ll m_D^{-1}$, which translates the fact that the maximum transverse separation between the quark and the antiquark is smaller than the Debye screening. The decoherence parameter (\ref{eq:amed}) can then be expanded for small dipole sizes $r_\perp$. The integral in eq.~(\ref{eq:amed})  becomes straightforward, yielding
\be
\label{eq:amed-dipole}
\left. \Delta_\text{med} \right|_{r_\perp^{-1} \gg m_D} = \frac{1}{6} \hat q L^{+} \,r_\perp^2 \left[\ln \frac{1}{ r_\perp m_D}+\text{const.}\right] \,.
\ee
This implies a coherent interaction of the antenna with medium and reveals the characteristic property of color transparency of a color singlet dipole.
\item {\it The ``saturation" regime}: where $r_\perp \gg m_D^{-1}$, i.e. the dipole size becomes larger  than the in-medium correlation length. The dipole cross section saturates in this case to a universal value that does not depend on the dipole parameters and reads
\be
\label{eq:amed-sat}
\left. \Delta_\text{med}  \right|_{r_\perp^{-1} \ll m_D} = \frac{ \hat q L^{+} }{m_D^2} \,.
\ee
Recalling that $\hat q \sim m_D^2/\lambda$, where $\lambda$ stands for the in-medium mean free path,
 we can rewrite (\ref{eq:amed-sat}) as $\Delta_\text{med} \approx L/\lambda$. This is nothing but the effective number of scattering centers which in this limit has to be smaller than or equal to one in order not to violate unitarity. In this case one needs to account for contributions from higher order in opacity. Indeed, it has been shown that $\Delta_\text{med}$ is bound by unity after including multiple scatterings \citep{MehtarTani:2011tz}. 
\end{itemize}
In what follows, we shall extend this picture to finite gluon energies.

\section{A problem of two scales}
\label{sec:scaling}

Going beyond the soft limit involves all terms in eq.~(\ref{eq:SpectrumMed}). The antenna radiation in medium is most clearly characterized in terms of the relevant transverse scales.\footnote{We would like to thank Al Mueller for inspiring this interpretation.} In transverse coordinate space, the two relevant scales in our case are the inverse Debye mass, $m_D^{-1}$, and the antenna size as probed by the medium, $r_\perp = \theta_\qqb L$, see eq.~(\ref{eq:rperp}). The inverse Debye mass characterizes the (electric) color screening of the medium, arising from its Yukawa-type interaction potential. It reflects the fact that the medium is {\it blanched} over transverse distances larger than $m_D^{-1}$.

\begin{figure}[t]
\centering
\includegraphics[width=0.45\textwidth]{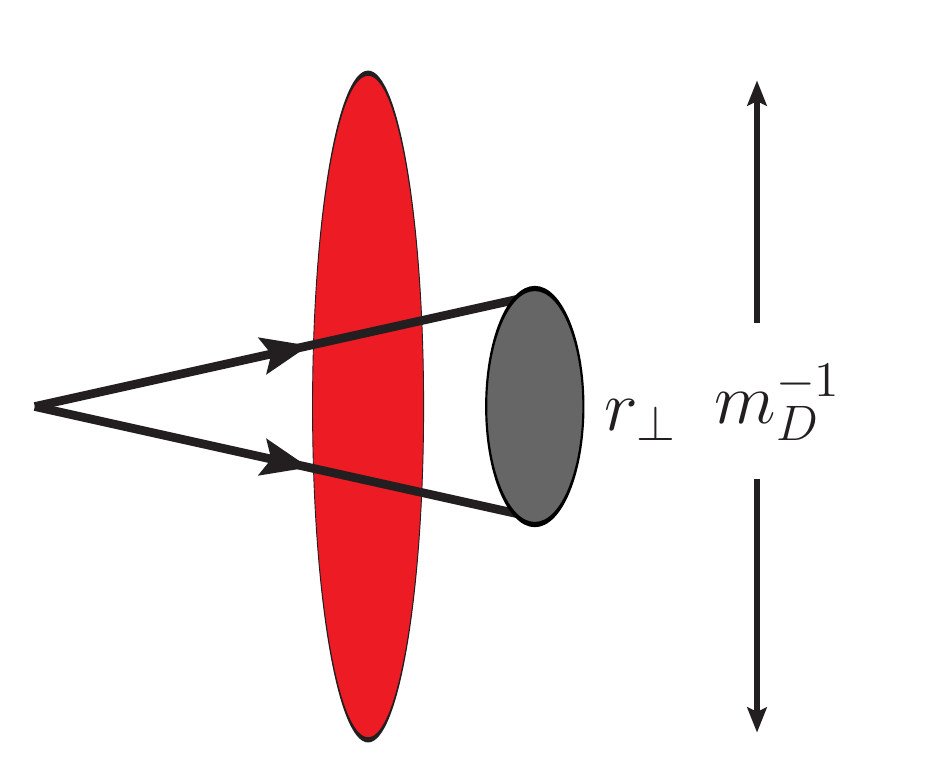}\hspace{1cm}
\includegraphics[width=0.38\textwidth]{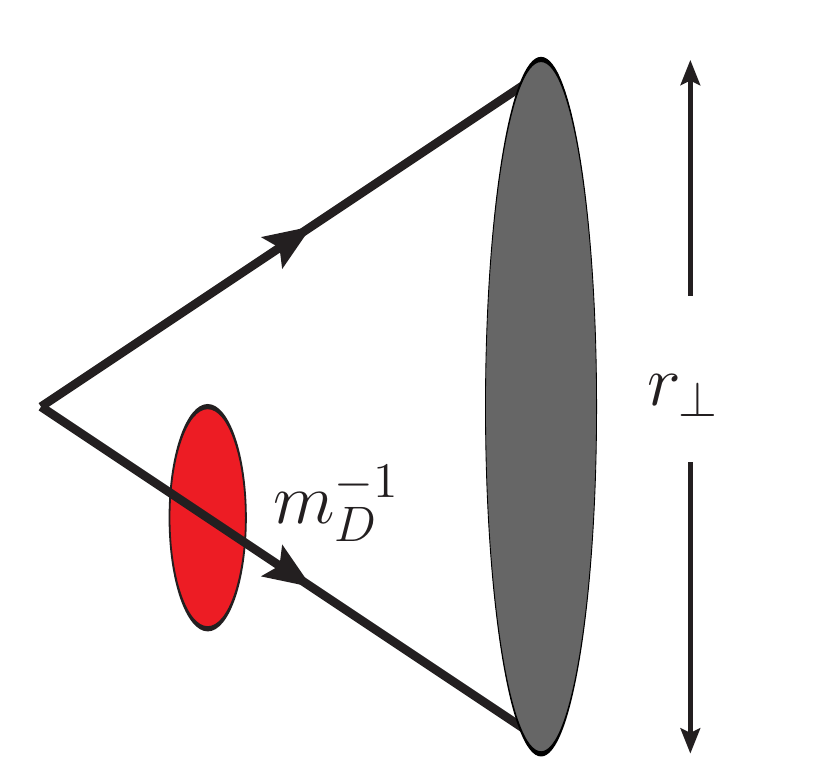}
\caption{The characteristic regimes of radiation in media: the ``dipole" regime, $r_\perp \ll m_D^{-1}$ (left) and the ``saturated" regime, $r_\perp \gg m_D^{-1}$ (right).}
\label{fig:regimes}
\end{figure}

The interplay between these scales is easily visualized in coordinate space, see fig.~\ref{fig:regimes}.  If the antenna is smaller than the transverse color screening length, $r_\perp \ll m_D^{-1}$, the medium can barely probe its inner structure and the cross-section is dominated by relatively hard medium modes $q_\perp\sim r_\perp ^{-1}$, see the left part of fig.~\ref{fig:regimes}. In this case, we expect the stimulated radiation to be dominantly coherent, i.e., excited off the antenna as a whole. Here, we expect the total spectrum not to depend much on the medium characteristics, namely $m_D$. This regime is therefore governed by the intrinsic scale of the antenna-dipole itself, namely $r_\perp^{-1}$, and we therefore call it the ``dipole" regime. 

In the opposite case, $r_\perp \gg m_D^{-1}$, cf. the right part of fig.~\ref{fig:regimes}, the medium cannot induce coherent emission off the antenna and thus the scale of the problem is set by the Debye mass. As already discussed in Section~\ref{sec:soft}, in this regime $\Delta_\text{med}$ has already saturated at its maximal value (proportional to the number of scattering centers $N_\text{scat}$) and we will therefore call it the ``saturation" regime.

This separation provides a clear and intuitive interpretation of the relevant physics and relates the geometric properties of an antenna (more generally, a jet), namely its transverse separation of correlated color, with the transverse color correlation of the medium. As will be demonstrated below, the physics of the two regimes is clearly separated and well-controlled.

Let us now go into more detail by analyzing the dynamics of the two regimes in the following subsections. In order to organize the discussion, we will first study the large-angle region when $\theta_\qqb$ is smaller than the typical angle of the independent medium-induced radiation $ \sim m_D/\omega$, i.e., where we expect interferences to be in force. This implies that this part of the discussion will be valid for $\omega < m_D/ \theta_\qqb$. Finally, we discuss the hard gluon sector, $\omega > m_D/\theta_\qqb$, before summarizing the analytic investigation.

\subsection{The ``dipole" regime}
\label{sec:dipoleregime}

This regime is characterized by $r_\perp^{-1} \gg m_D$, implying that the hardest momentum scale in the problem is set by the intrinsic antenna transverse momentum. 

As mentioned above, to begin with we focus on the range of gluon energies satisfying $\omega < m_D/\theta_\qqb$. The vectors $\Qh$ and $\Qhb$ simplify in this case to $\Qh\sim\Qhb\sim \Q$ which allows us to drop the last two terms in eq.~(\ref{eq:SpectrumMed}). The dominant contribution is then given by the first term in eq.~(\ref{eq:SpectrumMed}), such that the spectrum simply reads
\begin{multline}
\label{eq:SpectrumMed-dipoleregime}
\omega \left.\frac{dN^\text{med}}{d^3k}\right|_{r_\perp^{-1}\gg m_D}\!\! =\frac{8Ê\,\alpha_s C_F \,\hat q}{\pi} \potint \int_0^{L^+} \!\!dx^+  \left[1-\cos\left(\frac{\kh+\khb - 2\q}{2}\cdot\pip\, x^+\right)\right]  {\bs L}\cdot \bar{\bs L} \,.
\end{multline}
The discussion in this subsection will focus on how this term behaves with increasing $k_\perp$. With the approximations written above, eq.~(\ref{eq:SpectrumMed-dipoleregime}) can further be simplified to
\begin{multline}
\label{eq:SpectrumMed-dipregimaa}
\omega\frac{dN^\text{med}}{d^3k} \simeq \frac{8Ê\,\alpha_s C_F\, \hat q}{\pi}\ \potint \int_0^{L^+} dx^+  \Big\{1-\cos\left[({\bs k}-\q)\cdot\pip\, x^+\right]\Big\} \\
\times  \left[\frac{{\bs k}-\q}{({\bs k}-\q)^2}-\frac{\kh}{\kh^2}\right]\cdot \left[\frac{{\bs k}-\q}{({\bs k}-\q)^2}-\frac{\khb}{\khb^2}\right] \,.
\end{multline}
Note that as long as $k_\perp,q_\perp \lesssim r_\perp^{-1}$, the cosine can be expanded. Starting from the soft sector, where $k_\perp \ll m_D\lesssim q_\perp$, we get 
\beq
\label{eq:SpectrumMed-dipregimeb}
\omega\frac{dN^\text{med}}{d^3k}&\simeq&\frac{4Ê\,\alpha_s C_F \, \hat qL^+}{3\pi}\ \frac{\kh\cdot\khb}{\kh^2\khb^2}\potint \left(\q\cdot \r \right)^2 \,.\nn
&\simeq& \frac{\alpha_s C_F \, \hat q L^+}{3\pi^2}\frac{\kh\cdot\khb}{\kh^2\khb^2} \, r_\perp^2\left(\ln\frac{1}{r_\perp m_D}+\text{const.}\right) \,,
\eeq
where we have performed the $\q^2$ integral up to $r_\perp^{-2}$. This is nothing but the result obtained in the soft limit, cf. Section~\ref{sec:soft}, displaying the antiangular ordering feature, see eq.~(\ref{eq:lipatovsoft}). 

Now, continuing to angles larger than the opening angle of the pair, such that $\kh \sim \khb \sim {\bs k}$ and ${\bs k} \sim \q$, we get, 
\beq
\label{eq:SpectrumMed-dipregimec}
\omega\frac{dN^\text{med}}{d^3k} &\simeq& \frac{4Ê\, \alpha_s C_F\, \hat q L^+}{3\pi}\ \potint \left(({\bs k}-\q)\cdot\r\right)^2  \frac{\q^2}{({\bs k}-\q)^2{\bs k}^2} \nn
 & \simeq & \frac{\alpha_s C_F\, \hat q L^+}{3 \pi^2} \frac{r_\perp^2}{{\bs k}^2} \left(\ln \frac{1}{r_\perp m_D} + \text{const.} \right) \,.
\eeq
We observe the persisting logarithmic enhancement appearing above the characteristic medium scale, which verifies that it is not the relevant scale marking the onset of coherence. For large-angle radiation, on the other hand, i.e., $k_\perp> r_\perp^{-1}$, the cosine can be dropped, yielding simply
\beq
\label{eq:SpectrumMed-dipregimed}
\omega\frac{dN^\text{med}}{d^3k} &\simeq& \frac{8Ê\,\alpha_s C_F\, \hat q L^+}{\pi} \potint \frac{\q^2}{\Q^2{\bs k}^2} \nn
&\simeq& \frac{8Ê\,\alpha_s C_F\, \hat q L^+}{\pi} \frac{1}{{\bs k}^4} \ln\frac{{\bs k}^2}{m_D^2} \,,
\eeq
since already $k_\perp > q_\perp$. This spectrum is power-suppressed compared to the regimes described by eqs.~(\ref{eq:SpectrumMed-dipregimeb}) and (\ref{eq:SpectrumMed-dipregimec}), and demonstrates the cut-off of the spectrum above the characteristic scale, i.e, $k_\perp > r_\perp^{-1}$.

Now let us briefly comment on the case when $\omega>m_D/\theta_{q\bar q}$. This condition automatically implies that $\omega >\bar \omega_c$, since we are in the dipole regime. For these energies, the independent parts of the spectrum are strongly suppressed by the LPM effect, see Section~\ref{sec:glv}. Therefore, the last two terms in eq.~(\ref{eq:SpectrumMed}) can be neglected. The first term can be expanded around small $q_\perp$ leading back to eq. (\ref{eq:SpectrumMed-dipregimec}), which holds for $\theta>\theta_{q \bar q}$. For smaller angles, the spectrum is strongly suppressed. Thus, although the peak of the independent spectrum has entered the antenna cone, the antiangular ordering feature persists.

This situation implies in turn a ``maximal" energy for gluon emissions in this regime since the phase space for radiation vanishes when $\omega\theta_{q\bar q} > r^{-1}_{\perp}$. The interferences, and hence the total spectrum, is strongly suppressed for $\omega > \omega_d$, where $\omega_d \equiv (\theta_{q\bar q}^2 L)^{-1}$ (the subscript ``d" stands for dipole). Since $\omega_d > \omega_c$ in this regime, we expect an enhancement (``hardening") of the coherent spectrum compared to the independent one to persist up to this new characteristic energy, $\omega_d$.

In short, we have demonstrated that in the ``dipole" regime the spectrum is vacuum-like for angles larger than $\theta_\qqb$, i.e., $\sim 1/{\bs k}^2$, up to the characteristic scale $r_\perp^{-1}$. Beyond this scale the spectrum is power-suppressed. An important consequence of the fact that the medium scale, $m_D$, never plays a role in this regime is that the spectrum is not broadened, meaning that the spectrum is not enhanced around $k_\perp \sim m_D$.

\subsection{The ``saturation" regime}
\label{sec:saturationregime}

In this case the hardest scale is set by the medium, i.e., $r_\perp^{-1} \ll m_D$, and $\Delta_\text{med}$ is saturated at its maximal value. 

First, we restrict our discussion to relatively small gluon energies, $\omega < m_D/\theta_\qqb$. Since, in this regime $\bar \omega_c > m_D/\theta_\qqb$, this automatically implies that $\omega < \bar \omega_c$. These conditions imply, on one hand, $t_\text{form} < L$ and $(\kh+\khb -2\q )\cdot \pip L^+ > 1$, on the other. The arguments of both of the cosines in eq.~(\ref{eq:SpectrumMed}) are large, therefore we can drop them, and we simply get
\beq
\label{eq:SpectrumMed-2a}
\omega\left. \frac{dN^\text{med}}{d^3k}\right|_{r_\perp^{-1} \ll m_D} \!\!= \frac{4Ê\,\alpha_s C_F \, \hat q L^+}{\pi} \potint \left[ {\bs L}^2 + \bar {\bs L}^2 + \emcurr^2({\bs k}-\q) - \emcurr^2({\bs k})  \right] ,
\eeq
where ${\bs L}^2$ is given by eq.~(\ref{eq:lipatov}), $\emcurr^2({\bs k})$ by eq.~(\ref{eq:emcurr_vac}) and
\begin{align}
\label{eq:emcurr_shifted}
\emcurr^2({\bs k} -\q) &= \frac{(k^+ \pip)^2}{(\Qh)^2\, (\Qhb)^2} \,.
\end{align}
Before going into the detailed behavior of eq.~(\ref{eq:SpectrumMed-2a}), it is interesting to contrast the probabilistic interpretation of this expression to the independent radiation spectrum in (\ref{eq:GLV1}), as discussed after eq.~(\ref{eq:GLVprobabilistic}) \citep{Wiedemann:2000za}. For the antenna spectrum, we have a similar situation which only is affected by the presence of interferences. 

In other words, the situation when interferences between the hard emissions and the induced emissions, described by the Lipatov vertices, are negligible corresponds directly to the ``saturation" regime of the antenna. In this case, as in vacuum, these two mechanisms are clearly distinguishable. One piece describes the transverse momentum  re-shuffling (${\bs k} \to {\bs k} -\q$) of part of the bremsstrahlung gluons accompanying the creation of the hard antenna. This is given by the two latter terms of eq.~(\ref{eq:SpectrumMed-2a}), which both account for interference effects in like manner. This is in analogy with the two latter terms of eq.~(\ref{eq:GLVprobabilistic}), keeping in mind the absence of interferences in this expression.

In addition to the re-shuffling of hard, vacuum-like gluons, gluons are also induced independently off each of the antenna constituents, two first terms in eq.~(\ref{eq:SpectrumMed-2a}) which can be directly compared to the first term in eq.~(\ref{eq:GLVprobabilistic}) for the quark. By construction, these terms also drop rapidly at $k_\perp > m_D$. 

The detailed behavior of the spectrum follows the ideas sketched above keeping the energy constraint in mind. Now, in the soft limit $k_\perp \ll m_D$, we are left with the vacuum interference term, which reads
\beq
\label{eq:SpectrumMed-2b}
\omega\frac{dN^\text{med}}{d^3k} \simeq \frac{2Ê\,\alpha_s C_F}{\pi^2} \frac{\hat q L^+}{m_D^2}\frac{\kh\cdot\khb}{\kh^2\khb^2} \,,
\eeq
which is the antiangular ordered spectrum in the ``saturation" regime, see eq.~(\ref{eq:amed-sat}) in Section \ref{sec:soft}. For large angles, $\theta > \theta_{q\bar q}$, we can drop the dependence on the opening angle in all terms and finally we obtain
\begin{align}
\label{eq:SpectrumMed-2c}
\omega\frac{dN^\text{med}}{d^3k} &\simeq \frac{8Ê\,\alpha_s C_F \, \hat q L^+}{\pi}\ \potint \frac{\q^2}{{\bs k}^2\,\Q^2} \\
& \simeq \frac{2Ê\,\alpha_s C_F}{\pi^2} \frac{ \hat q L^+}{m_D^2} \left\{\begin{array}{lc} {\bs k}^{-2} & \quad k_\perp < m_D \\ m_D^2 \, {\bs k}^{-4} \ln \frac{{\bs k}^2}{m_D^2} & \quad k_\perp > m_D \end{array} \right. \,,
\end{align}
which again is the characteristic behavior of the Gunion-Bertsch spectrum. The cut-off of the spectrum is thus controlled by the typical medium scale $m_D$.

Let us presently turn to the hard part of the spectrum. For energies $\omega > m_D/\theta_\qqb$ the peak of the independent angular spectrum is localized at angles smaller than the opening angle of the pair. While interferences are already strongly suppressed for $\omega > m_D / \theta_\qqb$, cf. eq. (\ref{eq:SpectrumMed-2c}), the same is not true for the independent spectrum which will dominate in the energy interval $m_D/\theta_\qqb \lesssim \omega < \bar \omega_c$. Assuming that the opening angle is the largest angle in the problem, $m_D/\omega, \theta < \theta_\qqb$, the spectrum in the ``saturation" regime simplifies to 
\beq
\label{eq:insidethetaqq}
\left.\omega\frac{dN^\text{med}}{d^3k} \right|_{\omega \gg m_D/\theta_\qqb} \simeq \frac{8\,Ê\alpha_s C_F \, \hat q L^+}{\pi} \potint  \frac{{\bs k} \cdot \q}{{\bs k}^2 ({\bs k} -\q)^2} \,,
\eeq
where we have expanded around the quark contribution. Thus, for typical angles smaller than the opening angle and for energies $\omega > m_D/\theta_\qqb$ in the ``saturation" regime we recover the independent radiation spectrum (\ref{eq:GLV1}), taken in the same limit. In other words, after the peak of the independent distribution is well inside the cone, i.e., close to either of the constituents, the interferences are absent.

\subsection{Summary: general characteristics of the antenna spectrum}
\label{sec:summary}

The analysis of the various limits of the total antenna spectrum (\ref{eq:SpectrumMed}) in the two regimes can be summarized in a few lines. In Sections \ref{sec:dipoleregime} and \ref{sec:saturationregime} we have demonstrated that the spectrum for gluons with $\omega < m_D/\theta_\qqb$ is controlled by a single hard scale depending on the regime we are considering. This hard scale is simply
\beq
\label{eq:scales}
Q_\text{hard} \equiv \max \big( r_\perp^{-1}, m_D \big) \,,
\eeq 
for the ``dipole" and ``saturation" regimes, respectively. The transverse momentum of the gluon that can be induced in the system is naturally limited by $Q_\text{hard}$. The situation is also summed up in Table~\ref{tab:summary}, but let us here give an outline of the properties derived above. At small $k_\perp < \omega \theta_\qqb$ the medium-induced spectrum is suppressed due to the antiangular ordering property. As long as $\omega \theta_\qqb < k_\perp < Q_\text{hard}$ the transverse spectrum is leading and reads
\beq
\omega\frac{dN}{d\omega d{\bs k}^2} = \frac{2 \alpha_s C_F}{\pi} \frac{\Delta_\text{med}}{{\bs k}^2}  \,,
\eeq
and power-suppressed for larger momenta, $k_\perp > Q_\text{hard}$. For these energies, the average squared medium-induced transverse momentum is simply found to be
\beq
\langle k_\perp^2 \rangle_{\text{med}} \simeq \frac{Q^2_\text{hard}}{2} \ln^{-1} \frac{\omega_{\text{max}}}{\omega} \,,
\eeq
which holds for $\omega \ll \omega_{\text{max}}$, where 
\beq
\label{eq:omegamax}
\omega_{\text{max}} \equiv \frac{Q_\text{hard}}{\theta_\qqb} \,,
\eeq
is the maximum energy up to which interferences are suppressed, it arises due to the natural cut-off of the integral over the polar angle in the interferences in which $ \theta_\qqb < \theta < Q_\text{hard}/\omega$.

For $\omega > \omega_{\text{max}}$, the spectrum in the ``dipole" regime is vanishing. In the ``saturation" regime, though, the spectrum becomes dominated by the independent component for $m_D/\theta_\qqb<\omega < \omega_c$. Then, the average squared transverse momentum becomes $\langle k_\perp^2 \rangle = m_D^2 /2$ \citep{gyu00,Wiedemann:2000ez,Wiedemann:2000za}. 

\begin{figure}
\centering
\includegraphics[width=0.65\textwidth]{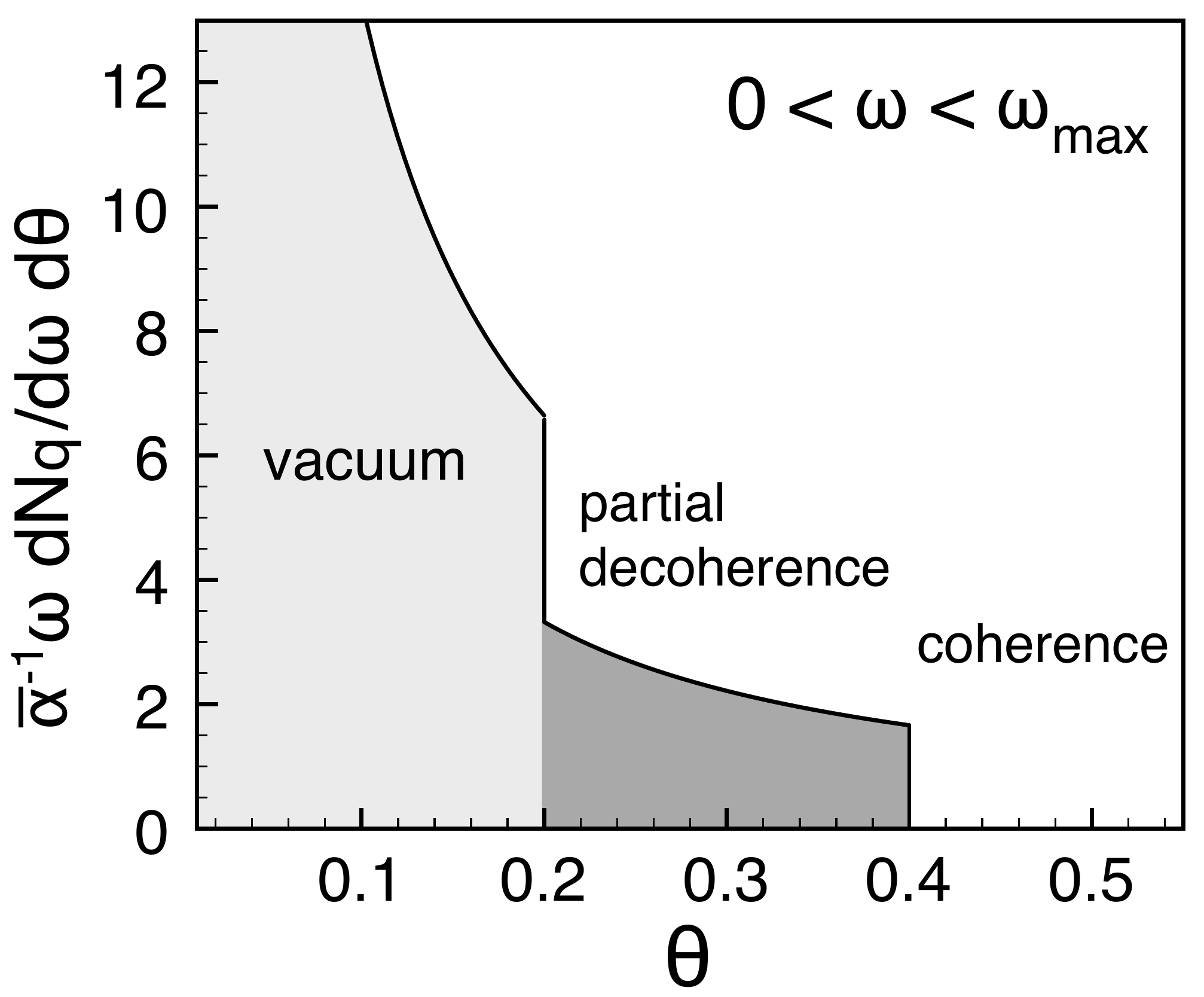}
\caption{Inclusive gluon production off one of the $\qqb$-antenna constituent, after azimuthal angle average, for energies $0 < \omega < \omega_\text{max}$. Parameters are $\theta_\qqb = 0.2$, $\Delta_\text{med} = 0.5$ and $Q_\text{hard}/\omega = 0.4$.}
\label{fig:VacMedIllustration_partial}
\end{figure}
For clarity, let us in parallel consider the behavior of the inclusive angular spectrum, i.e., after averaging over the azimuthal angle around the quark. The results obtained above confirm the discussion in Section~\ref{sec:soft}, namely that the spectrum is strongly suppressed at angles smaller than the opening angle, $\theta < \theta_\qqb$, and drops like $\sim 1/\theta$, weighted by the decoherence parameter $\Delta_{\text{med}}$, outside, $\theta > \theta_\qqb$. Turning to finite gluon energies, the appearance of the hard scale (\ref{eq:scales}) leads to an additional suppression of the medium-induced spectrum at large angles, namely for $\theta > Q_\text{hard}/\omega$, regardless of the regime under examination, where the coherent nature of emissions is restored. These features are illustrated in fig.~\ref{fig:VacMedIllustration_partial}. Such a characteristic angular cut-off is indeed seen in the numerics, cf. the two right panels of fig.~\ref{fig:theta_summary} below. 

Thus, the angular phase space for gluon emission, which in vacuum is determined by the angular ordering condition $\theta < \theta_\qqb$, is in medium augmented by a region of ``partial decoherence", i.e., modulated by the decoherence parameter $\Delta_\text{med}$, which extends up to the angle $Q_\text{hard}/\omega$. Above this angle full coherence is restored.

This are the general features encoded in the hard emission currents in eqs. (\ref{eq:emcurr_vac}) and (\ref{eq:emcurr_shifted}) and generalizes the picture of medium-induced decoherence described in Section \ref{sec:soft} and refs.~\citep{MehtarTani:2010ma,MehtarTani:2011tz}.

Let us wrap up the discussion by considering hard gluon emissions. While interferences are already strongly suppressed for $\omega > m_D/\theta_\qqb$ in the ``dipole regime" due to longitudinal interference effects (the LPM effect), the same is not true for the ``saturation" regime where the independent spectrum will dominate in the energy interval $m_D/\theta_\qqb \lesssim \omega < \bar \omega_c$. In this case, the antenna spectrum is predominantly the superposition of two independent spectra and the bulk of the independent radiation takes place at smaller angles than the opening angle, see Section \ref{sec:saturationregime}.

\section{Numerical results}
\label{sec:numerics}

We proceed with a numerical evaluation of the antenna spectrum. Following the strategy of Section~\ref{sec:vacuum}, we divide the spectrum into coherent contributions off the quark an the antiquark, namely 
\be
dN^\text{med}=dN_q^\text{med}+dN_{\bar q}^\text{med} \,,
\ee
where 
\beq
\label{eq:cohq}
\omega \frac{dN^\text{med}_q}{d^3 k} = \frac{\alpha_s C_F}{(2\pi)^2 \, \omega^2} \left(\mathcal{R}^\text{med}_q-\mathcal{J}^\text{med}_q\right) \,.
\eeq
The independent spectrum $\mathcal{R}^\text{med}_q$ was already discussed in detail in Section~\ref{sec:glv} and is defined explicitly in eq.~(\ref{eq:SpecGLV}). The interferences, on the other hand, are not as simply recovered as in the vacuum case. 

By looking at the phase structure in eq.~(\ref{eq:SpectrumMed}) it becomes clear that the product of Lipatov vertices in the first line of eq.~(\ref{eq:SpectrumMed}) comes with a phase related to the pair as a whole while the two remaining terms are dictated by the formation time of emissions off each of the components. Therefore, we divide the ${\bs L} \cdot \bar {\bs L}$ contribution between the two constituents and associate the remaining components which comes with the identical phase structure, e.g., as in (\ref{eq:GLV1}) for the quark, to either the quark or the antiquark. This procedure yields
\begin{multline}
\label{eq:interfq}
\mathcal{J}^\text{med}_q = -32\,\pi\,\hat q \potint ~\int_0^{L^+} dx^+ \left\{\frac{1}{2}\left[1-\cos\left(\frac{\kh+\khb -2\q}{2}\cdot\pip\, x^+\right)\right] {\bs L}\cdot \bar{\bs L} \right. \\ \left. - \left[1-\cos \frac{(\Qh)^2}{2k^+} x^+ \right] \frac{\Qhb}{(\Qhb)^2}\cdot {\bs L}\right\} \,,
\end{multline}
for the quark interference spectrum. A completely analogous expression holds for the antiquark interference spectrum $\mathcal{J}^\text{med}_{\bar q}$. The total medium-induced interferences are simply $\mathcal{J}^\text{med} = \mathcal{J}^\text{med}_q + \mathcal{J}^\text{med}_{\bar q}$. This decomposition is consistent with the soft limit, see Section~\ref{sec:soft}.

We go on to analyze the coherent spectrum off the quark in terms of $\mathcal{R}^\text{med}_q$ and $\mathcal{J}^\text{med}_q$, that we average over the azimuthal angle of the quark as usual. This procedure holds for analyzing inclusive quantities, which are the goal of the present work.

We choose the quark to move along the $z$-axis, thus $|{\bs p}| = 0$, while the antiquark moves in the $x-z$ plane, with transverse momentum given by $|\bar {\bs p}| = \bar E \sin\theta_\qqb$. Furthermore, in the numerics, the momentum exchange with the medium is always assumed to be in the $x-y$ plane, i.e., transverse to the quark. Strictly speaking, this is only valid in the approximation of small opening angles, $\theta_\qqb\ll1$, and small-angle gluon emissions, $\theta\ll1$, and is consistent with the high-energy approximation used throughout the paper. We normalize all curves by setting the number of scattering centers in the medium to unity, i.e. $\hat q L^+/m_D^2 = 1$.

\begin{figure}
\centering
\includegraphics[width=\textwidth]{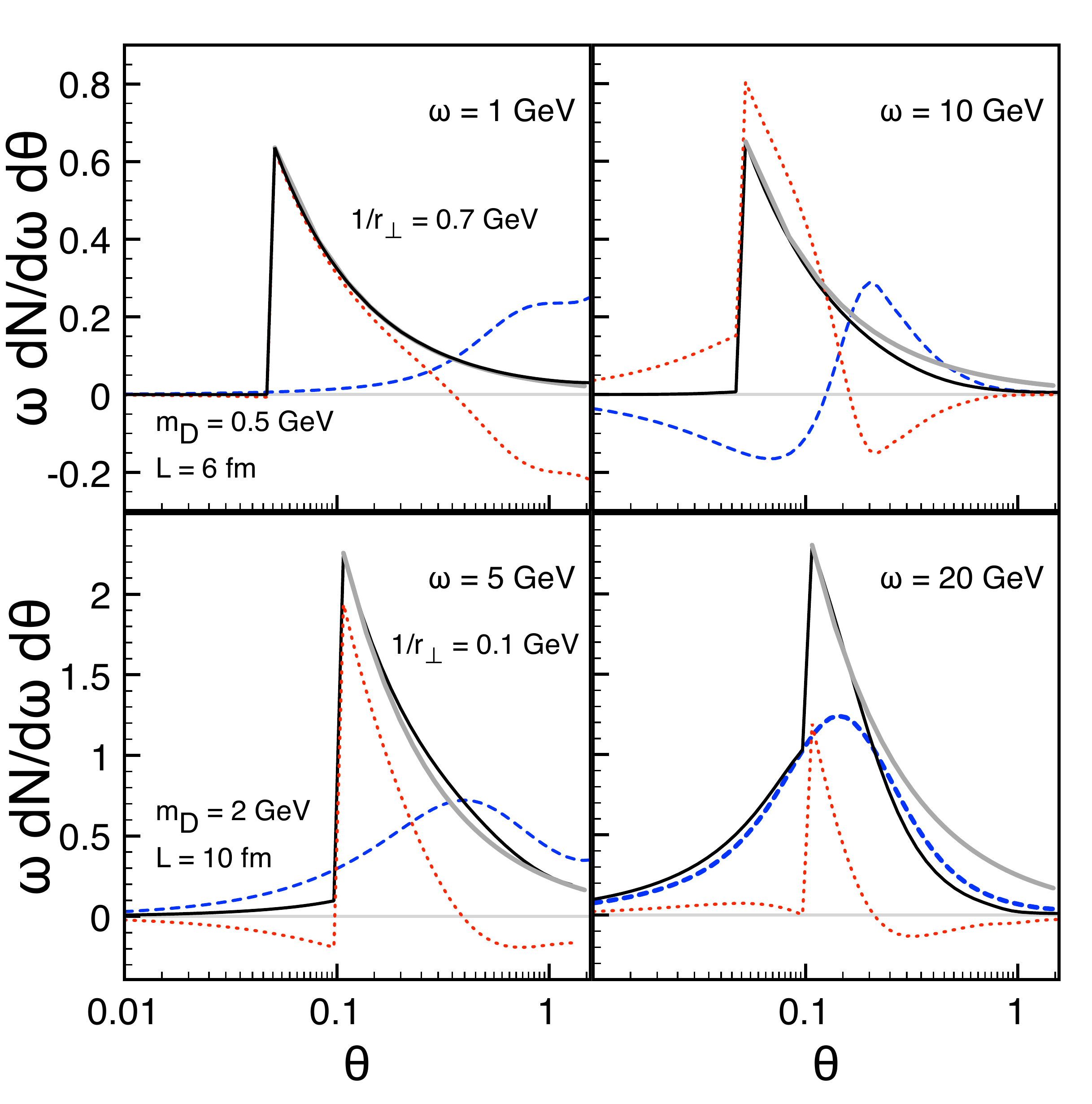}
\caption{The medium-induced angular spectrum. The solid (black) line corresponds to the total coherent spectrum off the quark (\ref{eq:cohq}), the dashed (blue) line is the independent spectrum (\ref{eq:SpecGLV}) and the dotted (red) line corresponds to the interferences (\ref{eq:interfq}). To guide the eye we have also depicted the vacuum-like antiangular ordered spectrum at large angles with a grey line ($\sim 1/\theta$). See text for further details.}
\label{fig:theta_summary}
\end{figure}
We show the coherent medium-induced spectrum off a quark, $\omega \, dN^\text{med}_q/d\omega d\theta$, in fig.~\ref{fig:theta_summary}. Parameters are chosen such that the two upper panels correspond to the ``dipole" regime while the two lower ones depict the typical situation in the ``saturation" regime. The independent spectrum is denoted with a dashed (blue) curve and the interferences with a dotted (red) curve. The solid (black) line represents the sum of both terms. To guide the eye, we have also depicted the vacuum-like antiangular ordered contribution ($\sim 1/\theta$, or $\sim 1/{\bs k}^2$) by a thick grey line. 

In the upper panels of fig.~\ref{fig:theta_summary}, the opening angle is small, $\theta_\qqb = 0.05$, and the medium parameters, $m_D = 0.5$ GeV and $L = 6$ fm, are chosen such that $r_\perp^{-1} \simeq 0.7$ GeV $>m_D$. We plot the results for two gluon energies: $\omega = 1$ GeV (upper, left panel) and $\omega = 10$ GeV (upper, right panel). In the lower two panels, parameters are chosen such that $r_\perp^{-1} < m_D$, namely $m_D = 2$ GeV, $L = 10$ fm and $\theta_\qqb = 0.1$ (so that $r_\perp^{-1} = 0.1$ GeV). The lower, left panel shows the result for $\omega = 5$ GeV, while $\omega = 20$ GeV in the lower, right panel.

In the soft limit, cf. upper, left panel of fig. \ref{fig:theta_summary}, the coherent spectrum adds up to zero at small angles, leaving the cone delimited by the pair angle free of radiation. The distribution jumps from zero inside the cone to a maximum value at $\theta=\theta_\qqb$, next it drops as $ 1/\theta$ for $\theta>\theta_{q\bar q}$. It differs notably from the independent component, see the dashed curve in the same panel, at all angles. This confirms the results of Section \ref{sec:soft}. 

Now turning to finite gluon energies, we see that radiation only takes place inside the opening angle of the $\qqb$-pair for the lower right panel, which represents the hard sector, $\omega > m_D/\theta_\qqb$, in the ``saturation" regime. As argued in Section \ref{sec:saturationregime}, this takes place because the peak of the independent spectrum starts entering the cone.

Furthermore, we have argued that regardless of the regime the medium-induced spectrum is suppressed for $k_\perp > Q_\text{hard}$. This is  observed in the angular distribution, cf. two right panels of fig.~\ref{fig:theta_summary}. The spectrum is clearly suppressed at large angles, namely for $\theta > Q_\text{hard}/\omega$, which shows to be relevant when $Q_\text{hard}/\omega < 1$ or, in other words, for gluon energies of the order of or larger than the hard scale, $Q_\text{hard}$. Initially, for $\theta \gtrsim \theta_\qqb$, the spectrum drops as the vacuum-like distribution, i.e., $\sim 1/\theta$. At some larger angle the spectrum changes behavior and drops faster. This is due to the power-suppression of the Gunion-Bertsch spectrum at large transverse momenta, $k_\perp \gg q_\perp$. Below, we will show that this cut-off is indeed controlled by the relevant hard scale in both the ``dipole" and ``saturation" regimes, see Sections \ref{sec:numericsdipole} and \ref{sec:numericssaturation}.

\begin{figure}
\centering
\includegraphics[width=\textwidth]{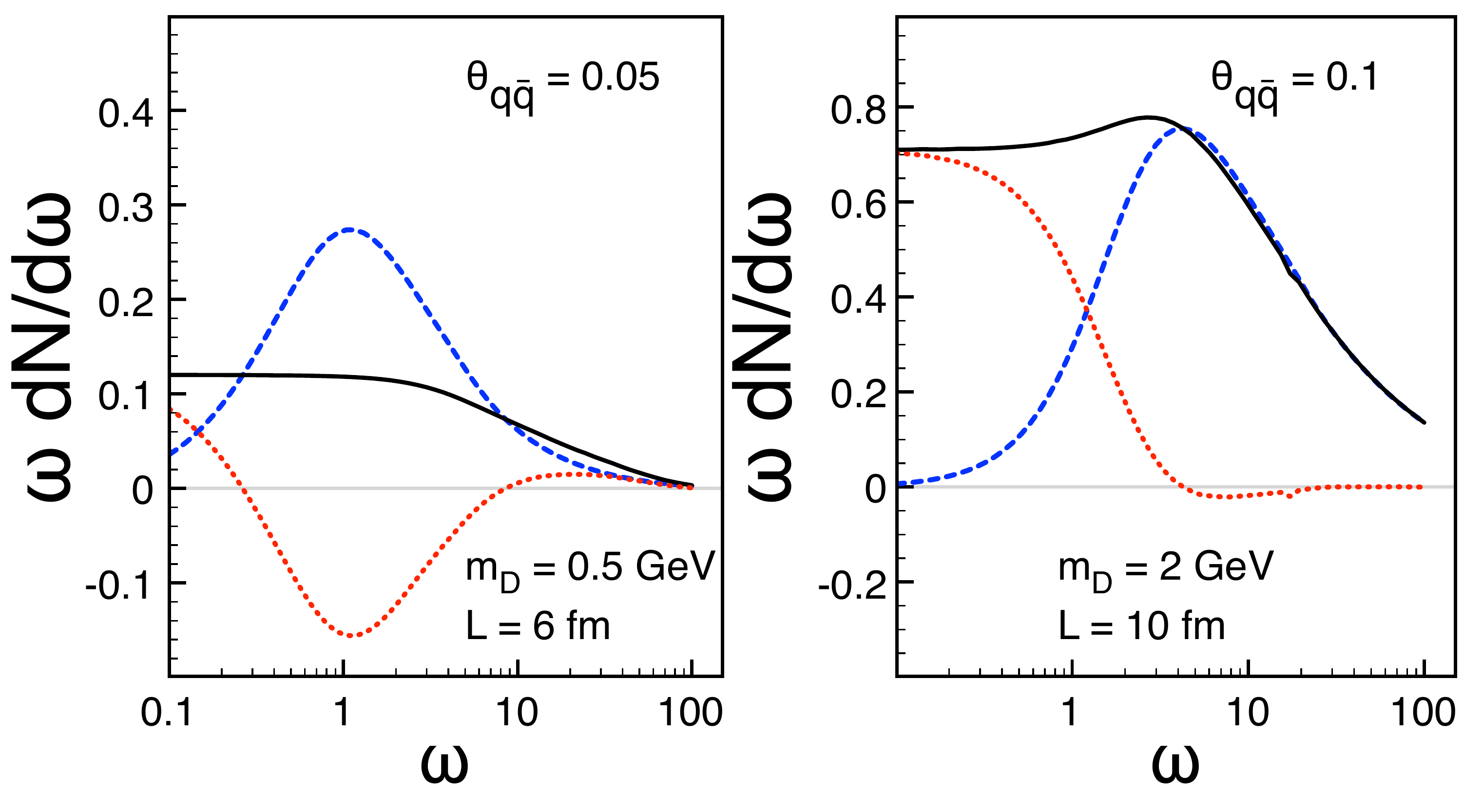}
\caption{The medium-induced gluon number distribution, given by (\ref{eq:SpectrumMed}) integrated over $\theta$, plotted for the two configurations in Fig~\ref{fig:theta_summary}. Notations are as in fig.~\ref{fig:theta_summary}.}
\label{fig:NumSpec}
\end{figure}
Next, we evaluate the medium-induced gluon energy spectrum
\beq
\omega \frac{dN^\text{med}_q}{d\omega} = \int_0^{\pi/2} \!\!d\theta ~\omega \frac{dN_q^\text{med}}{d\omega ~d\theta} \;,
\eeq
where the upper limit of the integral is chosen so that $k_\perp= \omega$, in fig.~\ref{fig:NumSpec}. The choice of parameters and notations are the same as previously, see fig.~\ref{fig:theta_summary}. The left panel in fig.~\ref{fig:NumSpec} therefore represents a typical energy distribution in the ``dipole regime, while the right panel gives an example for the ``saturation" regime.

In particular, the soft singularity present in the coherent spectrum, cf. (\ref{eq:lipatovsoft}), is apparent (in the soft limit, the spectrum is dominated by the interferences depicted by the dotted (red) curve). Such a contribution is absent in the independent, BDMPS-Z/GLV spectrum \citep{sal03}. For harder gluons, the spectra look quite different in the two regimes which follows the change of behavior of the interference contribution. In the ``dipole" regime, left panel of fig.~\ref{fig:NumSpec}, it cancels the independent spectrum up to large energies. What remains is a vacuum-like, logarithmic plateau up to the cut-off scale $\sim r_\perp^{-1}$. Since the hard scale in this regime is by definition larger than the typical scale governing the independent component, we observe a characteristic ``hardening" of the spectrum at large energies in the interval $m_D < \omega < r_\perp^{-1}$. In the ``saturation" regime, however, it is positive and extends the logarithmic plateau from the strictly soft sector to larger energies, i.e., up to the cut-off at $\omega \sim m_D$, see below. The spectrum of hard gluons, $\omega > m_D$, is in this case dominated by independent emissions and cleanly separated from the coherent sector, as expected.

These numerical results are in line with the discussion in Section \ref{sec:scaling}. In particular, they demonstrate how gluons with formation times smaller than the length, $t_\text{form} < L$, present in $\mathcal{J}^\text{med}_q$, contribute to the vacuum-like characteristics of the spectrum. What remains is to demonstrate the relevance of the hard scale $Q_\text{hard} = \max \left(r_\perp^{-1},m_D \right)$ for both of the regimes which appears as the scaling variable of the total spectrum.

\subsection{The ``dipole" regime}
\label{sec:numericsdipole}
\begin{figure}
\centering
\includegraphics[width=0.9\textwidth]{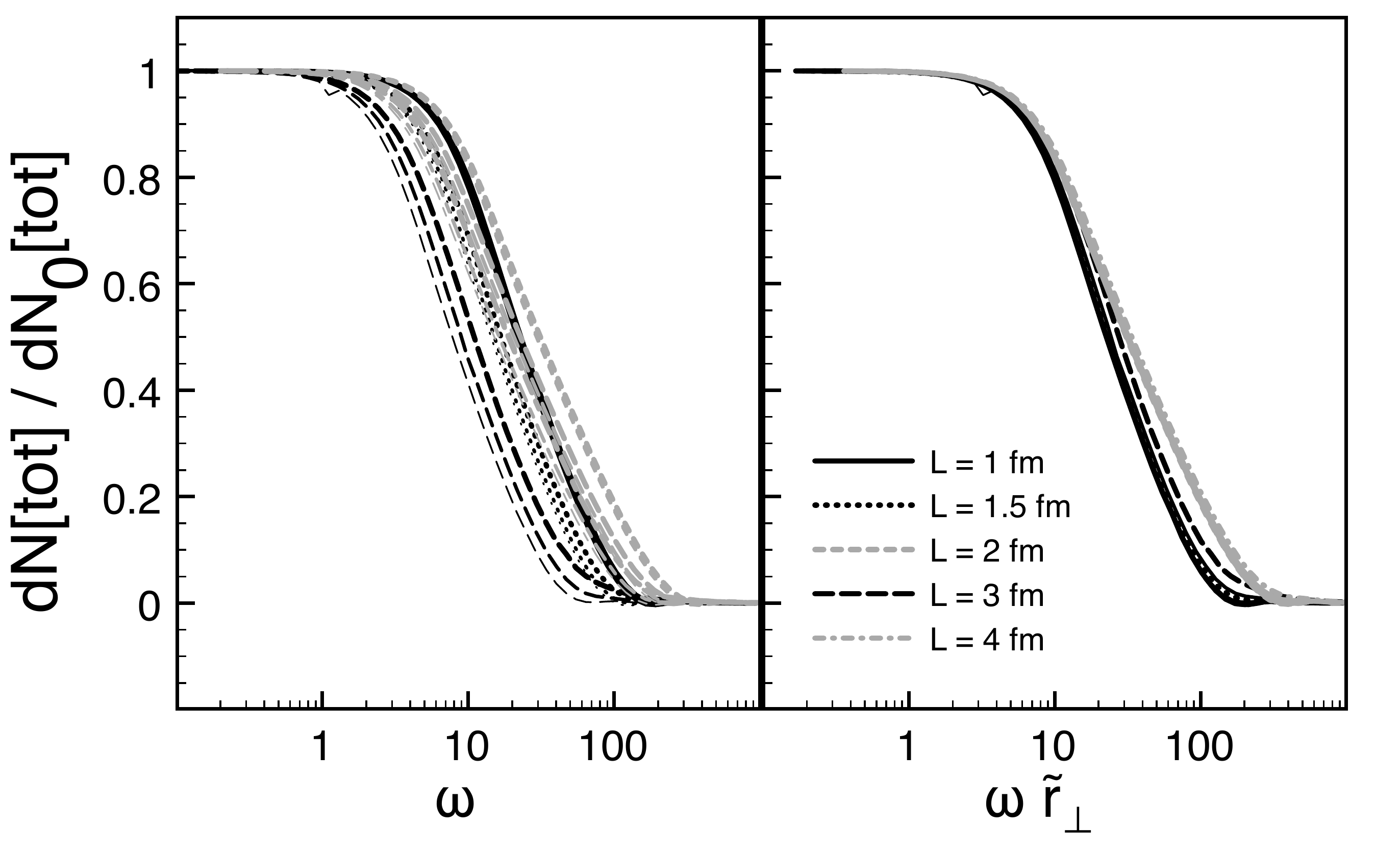}
\caption{The total medium-induced gluon energy spectrum in the ``dipole" regime. The spectra are scaled by their respective limiting values in the strictly soft limit. We have plotted curves for the following values of the parameters: $m_D =$ 0.8 (thick), 0.5 (medium) and 0.3 (thin curves); $L =$ 1 (solid), 1.5 (dotted), 2 (short-dashed), 3 (long-dashed) and 4 fm (dash-dotted curves); $\theta_\qqb = $ 0.1 (black) and 0.05 (grey curves). In the right panel of the figure, all curves are scaled with $\tilde r_\perp^{-1}$, see text for further discussion.}
\label{fig:ScalingDipole}
\end{figure}
This regime has previously not been investigated in the literature in the context of jet physics in medium. Let us quickly recall the main features of the angular spectrum in this case. The main trends found analyticall are confirmed in the upper panels of fig.~\ref{fig:NumSpec}: the spectrum is suppressed inside the cone, and behaves vacuum-like, $\sim 1/\theta$, up to a limiting angle where it is strongly suppressed. What remains to show is that this angle follows the scaling behavior $Q_\text{hard}/\omega$, where $Q_\text{hard} = r_\perp^{-1}$ in this case. This cut-off becomes manifest when we integrate over the angle, leading to a characteristic cut-off of the energy spectrum which scales simply as $r_\perp^{-1}$.

To prove that this is the case, we have plotted the medium-induced energy spectrum for a set of parameters which all respect $r_\perp^{-1} > m_D$ in fig.~\ref{fig:ScalingDipole}, for details see the figure caption. The spectra have been scaled by their respective limiting values in the strictly soft limit to ease the comparison. In the right panel of fig.~\ref{fig:ScalingDipole}, we have scaled the energies by $\sim r_\perp^{-1}$. As a matter of fact, due to the hard, Coulomb-like tail of the medium potential (\ref{eq:MediumPotential}) a logarithm should appear together with $r_\perp$, e.g. see the dipole cross section (\ref{eq:DipoleCrossSection}) and the discussion there. To ensure that the logarithm is slowly varying, as it should be in the ``dipole" regime, in fig.~\ref{fig:ScalingDipole} we have explicitly scaled the energy with $\tilde r_\perp = r_\perp \sqrt{\ln 20/(r_\perp m_D)}$. Keeping in mind this minor complication, it is apparent that the characteristic cut-off scales according to the hard scale. The discrepancy for curves with different opening angles at even larger gluon energies, $\omega > m_D$, comes about due to the cut-off on the angular integral, $\omega_{\text{d}}$ in (\ref{eq:omegamax}), which grows with decreasing $\theta_\qqb$.

The total spectrum in the ``dipole" regime is therefore not sensitive to the medium characteristics, except for the slow, logarithmic dependence due to the tail of the medium potential, and is simply governed by the intrinsic momentum scale of the antenna-dipole. 

\subsection{The ``saturation" regime}
\label{sec:numericssaturation}
\begin{figure}
\centering
\includegraphics[width=0.9\textwidth]{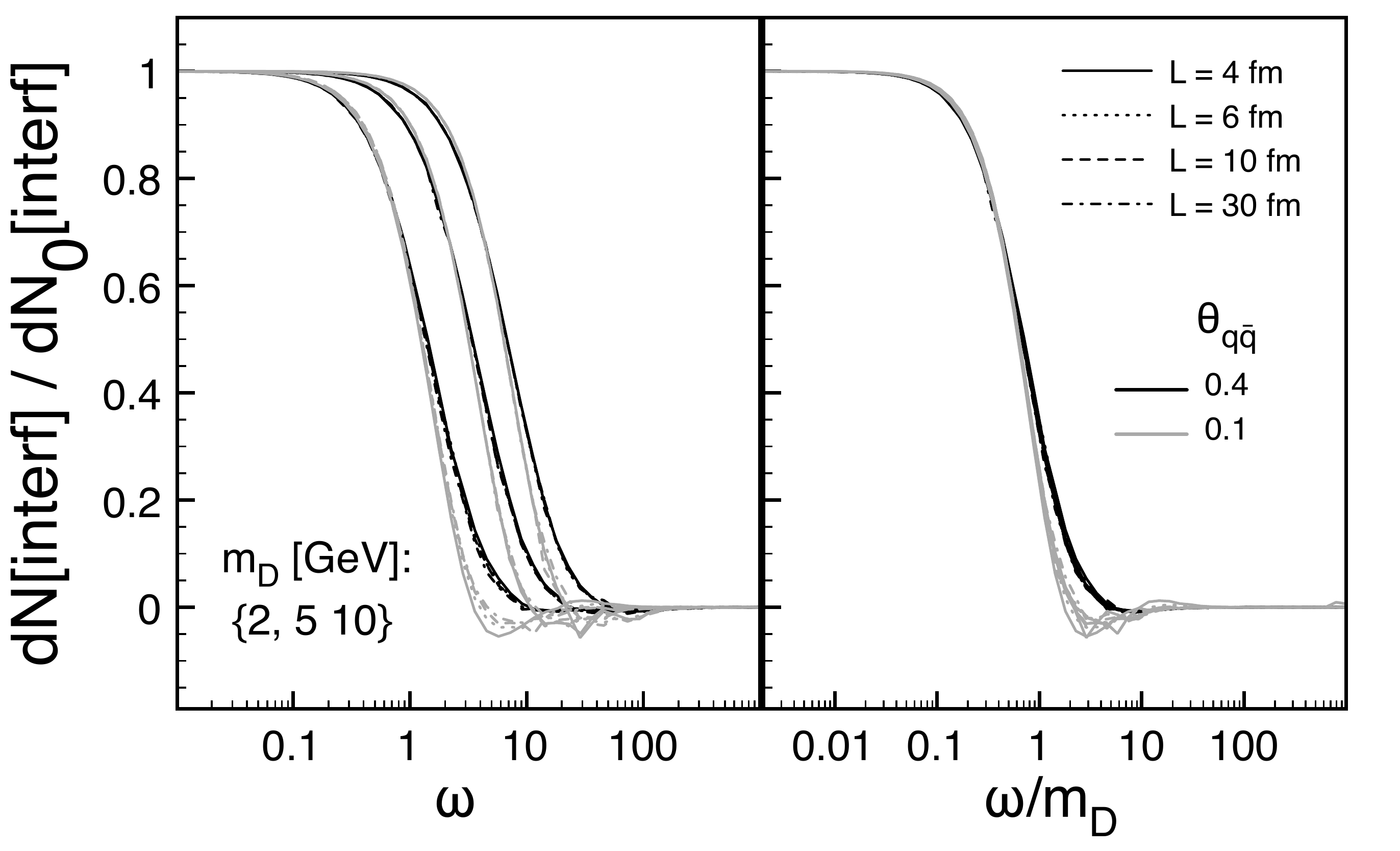}
\caption{The medium-induced interference contribution, $\mathcal{J}_q^\text{med}$, to the gluon energy spectrum in the ``saturation" regime. The spectra are scaled by their respective limiting values in the soft limit. Black curves are for opening angle $\theta_\qqb = 0.4$ and grey curves for $\theta_\qqb = 0.1$, and we have evaluated the spectrum for lengths $L =$ 4, 6, 10 and 30 fm. In the left panel of the figure, the curves group according to the value of $m_D$ which takes the values: 2, 5 and 10 GeV, starting from the left. Curves for different $L$ are overlapping. In the right panel, all curves are scaled with $m_D$.}
\label{fig:ScalingDense_J}
\end{figure}
In this regime, the only relevant scale should be provided by the characteristic transverse momentum transferred from the medium, simply $m_D$ in our case. As in the previous case, we observe the characteristic angular cut-off at $\theta_\text{max}$, cf. bottom, right panel of fig.~\ref{fig:theta_summary}, which in this case is given by $\theta_\text{max} = m_D/\omega$. 

In particular, it is well-known that the peak of the independent energy distribution is positioned at a characteristic energy $\sim m_D$ \citep{sal03}, see fig.~\ref{fig:GLV} and the discussion in Section~\ref{sec:glv}. What remains to show is that the interferences, $\mathcal{J}_q^\text{med}$, are scaling likewise. This is done in fig.~\ref{fig:ScalingDense_J}, where we plot the medium-induced interferences for the quark (\ref{eq:interfq}) for a range of parameters such that $r_\perp^{-1} < m_D$, for details see the figure caption. The curves are once again scaled by their respective values in the strictly soft limit so as to ease the comparison. In the right panel, we have scaled the energy by $m_D$ and, as expected, a nearly perfect scaling is obtained.

We conclude that in the ``saturation" regime the intrinsic momentum scale of the antenna, $r_\perp^{-1}$, does not play a significant role. Nevertheless, as seen for the curves with large $m_D$ in the left part of fig.~\ref{fig:ScalingDense_J}, and also in the right panel of fig.~\ref{fig:NumSpec}, the contribution from the interferences dominate the soft sector and extend to semi-soft values, i.e., $\omega \lesssim m_D$. Above the characteristic medium momentum scale, though, the medium-induced interferences are negligible and one is only left with independent emissions.

\section{Antenna radiation in medium -- the color octet case}
\label{sec:octet}

Above we have discussed the medium-induced radiation off a $\qqb$ pair which is produced in a color singlet configuration, e.g., originating from the splitting of a virtual photon. In heavy-ion collisions, a physically more relevant process would for instance be that of gluon fragmentation in the presence of a medium. 

The choice of setup for the calculations above was motivated by the simplicity of the color algebra. Actually, the calculation for other color configurations follows exactly the same steps as in Section~\ref{sec:medium}, and taking care of the non-zero total charge of the system. The virtuality of the initial gluon, assumed to be the largest scale of the problem, provides the factorization of the splitting process and the subsequent radiation. Thus, both emission off and medium-interaction of the initial gluon are suppressed by the large scale and can safely be neglected in the leading logarithm approximation. 

As an explicative example, we compute the spectrum for the splitting of a highly virtual gluon into an quark-antiquark antenna, $g^\ast \to \qqb$. In this case the medium-induced spectrum simply becomes
\begin{equation}
\label{eq:SpectrumMedOctet}
\omega\frac{dN^\text{med}}{d^3 k} = \frac{\alpha_s}{(2\pi)^2 \omega^2} \Big[ C_F\left( \mathcal{R}_q^\text{med} + \mathcal{R}_{\bar q}^\text{med}- \mathcal{J}^\text{med}\right) + \frac{C_A}{2} \mathcal{J}^\text{med} \Big] \,,
\end{equation}
where $\mathcal{J}^\text{med} = \mathcal{J}^\text{med}_q + \mathcal{J}^\text{med}_{\bar q}$. The singlet spectrum, see eq.~(\ref{eq:SpectrumMed}), is recovered by putting $C_A$ to 0. The latter term in eq.~(\ref{eq:SpectrumMedOctet}) is interpreted as the medium-induced spectrum off the total charge, i.e., the parent gluon {\it imagined} to be on-shell, in analogy to the octet antenna radiation in vacuum, see eq.~(\ref{eq:spec-general}). This becomes more apparent when considering the limit $\theta_\qqb \to 0$, when the medium-induced singlet component of the spectrum vanishes owing to the conservation of color. What remains, is
\begin{equation}
\left. \omega\frac{dN^\text{med}}{d^3 k} \right|_{\theta_\qqb \to 0} \simeq \frac{\alpha_s C_A}{(2\pi)^2 \omega^2} \mathcal{R}^\text{med} \,,
\label{eq:spec-medium-octet} 
\end{equation}
which is nothing but the medium-induced radiation spectrum off the parent gluon.

Adding the vacuum contribution and restricting ourselves to the soft gluon emissions the total radiation spectrum reads
\begin{equation}
\omega\frac{dN}{d^3 k}\Big|_{\omega\to0} = \frac{\alpha_s}{(2\pi)^2 \omega^2}\left[C_F\left(\mathcal{R}_q+\mathcal{R}_{\bar q}-2\left(1-\Delta_{\rm med}\right)\mathcal{J}\right)+C_A(1-\Delta_{\rm med})\mathcal{J}\right],
\end{equation}
which proves again that the role of the medium-induced emissions is to break the coherence of the quark and the antiquark: the terms proportional to $C_F$, the same as in the singlet case, determine the onset of decoherence by medium-induced antiangular ordered emission; the term proportional to $C_A$ shows a reduction of the large-angle radiation off the total charge. The latter feature is reminiscent of a ``memory loss" effect \citep{MehtarTani:2011tz}. This is in line with the breakdown of angular ordering and decoherence for soft gluon emissions in the medium.

\section{Conclusions and perspectives}
\label{sec:conclusions}

One of the most urgent questions in the theory and phenomenology of nuclear collisions at high energies is how a consistent treatment of the in-medium parton shower should be formulated. The standard approach, extensively studied at RHIC, is based on the energy-losses  due to  the medium-induced gluon radiation off a highly energetic quark or gluon traversing a medium. This approach, essentially valid for computing the medium effects in high-$p_T$ inclusive particle production, is not adequate for more differential observables such as reconstructed jet substructure. For such a class of observables the whole parton shower needs to be accounted for and, in particular, the presence of ordering variables, encoded in evolution equations, unraveled. Following the logic and insight gained from the vacuum case, the antenna constitutes a very convenient laboratory, where most of the needed ingredients are already present. In this simplified setup, the gluon radiation off a pair of partons, correlated in color at the moment of creation, is studied. In the vacuum, this allows the identification of angular ordering of the radiation which follows from color interference effects among the  emitters. A consistent treatment for inclusive quantities of the subsequent emissions reveals that the parton shower evolution can be reduced to the problem of antenna radiation with ever decreasing opening angles. These features are encoded, in particular, in Monte Carlo codes.

The question is whether these coherence effects survive the passage through the medium. To study this problem, and with the hope of finding a first step towards a theoretically motivated in-medium parton shower, the medium-induced spectrum off a $\qqb$ antenna was first calculated in \citep{MehtarTani:2010ma} at first order in opacity. Most importantly, in the limit of soft gluon emissions we obtained two novel features compared to previous calculations of medium-induced radiation. Firstly, the spectrum was divergent in the limit of vanishing gluon energies. Secondly, a geometrical separation between vacuum and medium-induced radiation, called antiangular ordering, was obtained. In this work we have detailed the analytical and numerical calculations underlying the results in ref.~\citep{MehtarTani:2010ma} and extended the corresponding discussion. The main result is the total spectrum in eq.~(\ref{eq:SpectrumMed}).

We have found that the behavior of the total spectrum is determined by the hardest scale of the problem, i.e., $Q_\text{hard}\equiv \max \left(r_\perp^{-1}, m_D \right)$. We have called these separated kinematical regimes the ``dipole" and ``saturation" regimes, respectively. Let us shortly summarize the main features for each of them, referring to Section \ref{sec:scaling} for a complete discussion.

The radiation in the ``dipole" regime is mainly induced at angles larger than the opening angle of the pair, $\theta > \theta_\qqb$. The intensity of the radiation is proportional to the decoherence parameter $\Delta_\text{med} \propto r_\perp^2$ and behaves vacuum-like up to angles $(r_\perp \omega)^{-1}$, see fig.~\ref{fig:VacMedIllustration_partial}. At higher angles the spectrum is strongly suppressed.

The behavior of the radiation in the ``saturation" regime can roughly be separated into two parts. In this case, it is the medium characteristics, $m_D$, that sets the hard scale. For energies $\omega \lesssim m_D/\theta_\qqb$, the spectrum behaves as in the ``dipole" regime except for two key aspects: the decoherence parameter is constant (saturated) and the maximal angle of radiation is now given by $m_D/\omega$, see fig.~\ref{fig:VacMedIllustration_partial}. For hard gluons, $\omega > m_D/\theta_\qqb$, the medium-induced spectrum is dominated by radiation that is located at angles smaller than the opening angle, emitted independently by each of the constituents of the antenna. 

The results presented in this paper are truncated at first order in medium opacity and, thus, do not include unitarity constraints which become imperative when the number of scattering centers in the medium is large. A consistent treatment should in such a situation also account for interference effects among multiple scattering centers (the LPM effect). In the limit of very opaque media, the medium decoherence parameter saturates at the maximal value given by unitarity, $\Delta_\text{med} \to 1$. This marks the onset of a ``dense" regime, which we plan to study in an upcoming publication. The antenna radiation in medium was studied for the multiple-soft scattering approximation in \citep{MehtarTani:2011tz} for the soft limit and approached analytically in \citep{MehtarTani:2011jw,CasalderreySolana:2011rz}. 

\section*{Acknowledgments}

The authors would like to thank N.~Armesto, Yu.~Dokshitzer, H.~Ma, M.~Martinez and A.~H.~Mueller for helpful discussions. This work is supported by Ministerio de Ciencia e Innovaci\'on of Spain; by the Spanish Consolider-Ingenio 2010 Programme CPAN and in part by the Swedish Research Council (contract number 621-2010-3326). CAS is a Ram\'on y Cajal researcher.

\appendix
\section{The Feynman diagram calculation}
\label{sec:feynman}
As a further cross-check of the calculation within the classical Yang-Mills (CYM) formalism, presented in Section~\ref{sec:medium}, we have in parallel also calculated the process under consideration using the standard Feynman diagram technique. Below we describe how to obtain the relevant cross section at first order in opacity for an antenna in a singlet state. The calculation of the octet case involves only a more complicated color structure. The cross section factorizes into two factors describing the the elastic splitting process, $\gamma^\ast \to \qqb$, and the subsequent radiation off the quark or antiquark. Thus, apart from the former part, the radiative cross section is simply
\beq
\label{eq:FeynCrossSection}
d\sigma_{(1)} = d\sigma_{(0)} \left\{ \langle\left| \mathcal{M}_{(1)} \right|^2 \rangle + 2 \,\text{Re}\langle \mathcal{M}_ {(2)} \mathcal{M}_{(0)}^\ast \rangle \right\} \frac{d^3k}{(2\pi)^3\, 2\omega}\;,
\eeq
where $d\sigma_{(0)}$ is the Born-level cross section. The first term in eq.~(\ref{eq:FeynCrossSection}), $\mathcal{M}_{(1)}$, is the sum of all diagrams involving one scattering center and the second, $\mathcal{M}_{(2)}$, the corresponding sum for two scattering centers in the contact limit. It accounts for the fact that we do not measure (tag) the interacting medium gluons and thus have to take into account virtual corrections to the interaction (so-called contact terms) not to violate unitarity. The shorthand $\langle \dotsm \rangle$ in eq.~(\ref{eq:FeynCrossSection}) denotes medium averages, which are given by (\ref{eq:MediumAverage}) for the single-scattering diagrams and (\ref{eq:MediumAverage2}) for the contact terms.

All calculations are done in the high-energy (eikonal) approximation. As for the previous calculations, we will assume massless quarks and work in light-cone gauge, defined by the condition $A^+ =0$. The kinematics are given in Section~\ref{sec:vacuum}.

\subsection{$N=0$ vacuum amplitudes}
To set the conventions for the Feynman rules used in the following, we list the amplitudes for antenna radiation in vacuum. With the gauge conditions described above and assuming the gluon emission vertex to be eikonal (no recoil), they are given by
\beq
\mathcal{M}_{(0)1}^a = \begin{minipage}{1cm} \includegraphics[width=\textwidth]{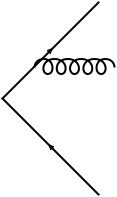} \end{minipage} &\simeq& -2g\, \mathcal{M}_0(p,\bar p)\,  t^a \, \frac{\kh \cdot \epsp}{\kh^2} e^{i k\cdot x_0}\\
\mathcal{M}_{(0)2}^a = \begin{minipage}{1cm} \reflectbox{\includegraphics[angle=-180,width=\textwidth]{GluonAmp.png}} \end{minipage} &\simeq&2 g\,\mathcal{M}_0(p,\bar p)\,  t^a \, \frac{\khb \cdot \epsp}{\kh^2} e^{i k\cdot x_0}
\eeq
where $\mathcal{M}_0 (p,\bar p) \equiv \bar u(p) e^{i p\cdot x_0}\,\mathcal{H}(x_0)\, e^{i \bar p \cdot x_0} v(\bar p)$ and $x_0$ denotes the point where the $\qqb$ pair was produced. Here $\mathcal{H}(x_0)$ denotes the hard production vertex that factors out for soft gluon radiation. 

\subsection{$N=1$ amplitudes}
\label{sec:feynman-n1}
We go on to calculate the amplitudes with one medium interaction. The medium is modeled by a static potential at position $x_1$, described by $\A_\text{med}(x_1,\q)$ and denoted as a cross in the Feynman diagrams below, carrying an incoming 4-momentum $q$. From now on, we set $x_{0\perp}=x^-_0 = 0$. Then,
\begin{multline}
\mathcal{M}_{(1)1}^a = \begin{minipage}{1.2cm} \includegraphics[width=\textwidth]{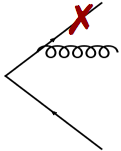} \end{minipage} 
 \simeq  -2i\, g^2 \,\mathcal{M}_0 (p,\bar p)  \, t^bt^a \, \int \!\! dx_1^+  \int \frac{d^2 \q}{(2 \pi)^2}\ \frac{\kh\cdot \epsp}{\kh^2} e^{ik^- x_0^+}  \\
 \times \A^{b}_\text{med}(x_1^+,\q) \ \left( 1-e^{i \frac{\kh^2}{2k^+}\Delta_{10}^+}\right)e^{i \frac{\pp \cdot \q}{p^+}\Delta_{10}^+ } \;,
\end{multline}
where we have defined the longitudinal separation $\Delta_{10}^+ = x_1^+-x_0^+$. Similarly,
\begin{multline}
\mathcal{M}_{(1)2}^a = \begin{minipage}{1.2cm} \includegraphics[width=\textwidth]{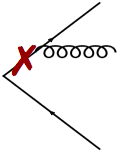} \end{minipage} \simeq -2i\, g^2 \,\mathcal{M}_0 (p,\bar p)\, t^a t^b \,\int \!\! dx_1^+ \int  \frac{d^2 \q}{(2 \pi)^2}\ \frac{\kh \cdot \epsp}{\kh^2}~e^{i k^- x_0^+} \\
\times \A^{b}_\text{med}(x_1^+,\q) \ e^{i \frac{\kh^2}{2k^+}\Delta_{10}^+}e^{i \frac{\pp \cdot \q}{p^+} \Delta_{10}^+ } 
\end{multline}
\begin{multline}
\mathcal{M}_{(1)3}^a = \begin{minipage}{1.2cm} \includegraphics[width=\textwidth]{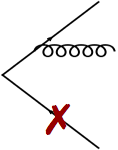} \end{minipage} \simeq 2 i\, g^2 \,\mathcal{M}_0(p,\bar p)\, t^a t^b \, \int\!\! dx_1^+ \int \frac{d^2 \q}{(2 \pi)^2} \ \frac{\kh \cdot \epsp}{\kh^2}\, e^{i k^- x_0^+} \\
\times  \A^{b}_\text{med}(x_1^+,\q) \ e^{i \frac{\ppb \cdot \q}{\bar p^+}\Delta_{10}^+}\,.
\end{multline}
Next, in the case of gluon interaction the triple gluon vertex is given by
\begin{multline}
\mathcal{M}_{(1)4}^a = \begin{minipage}{1.2cm} \includegraphics[width=\textwidth]{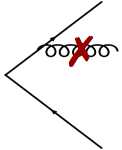} \end{minipage} \simeq 2 i\, g^2 \, \mathcal{M}_0(p,\bar p)\, [t^a,t^b] \,\int \!\! dx_1^+  \int \frac{d^2 \q}{(2\pi)^2} \frac{(\Qh) \cdot \epsp}{(\Qh)^2}  e^{ik^-  x_0^+}   \\
\times \A^{b}_\text{med}(x_1^+,\q)\, e^{i \frac{\kh^2}{2k^+}\Delta_{10}^+} e^{i  \frac{\pp \cdot \q}{p^+}\Delta_{10}^+}  \left( 1- e^{-i\frac{(\Qh)^2}{2k^+}\Delta_{10}^+}\right)\;.
\end{multline}
The corresponding diagrams for emission off the antiquark follow from symmetry.

\subsection{$N=2$ amplitudes in the contact limit}
\label{sec:feynman-n2}
In order to conserve unitarity of the inclusive cross section at first order in opacity, we have to take into account the double-interaction amplitudes in the contact limit. In other words, the color and the momentum transfer from the medium is conserved at the level of the amplitude. Furthermore, the interactions take place at the same longitudinal point on the light-cone. These terms thus serve as virtual corrections that regulate the total cross section in the limit $\q\rightarrow 0$. A subset of the double-interaction terms are usually called contact terms, namely the ones where the same particle interacts twice with the same scattering center.  

For these diagrams, we perform the medium averages on the level of the amplitude. Since both interactions takes place on the same side of the cut, we make use of a the fact that the medium background field is real, i.e, $\A_{\text{med}}^\ast(x^+,\q)=\A_{\text{med}}^\ast(x^+,-\q)$ in eq.~(\ref{eq:MediumAverage}), namely
\beq
\label{eq:MediumAverage2}
\langle \A^{a}_\text{med}(x^+,\q) \A^{b}_\text{med}(x'^+,\q')\rangle = \delta^{ab} m_D^2 n(x^+)\,\delta(x^+-x'^+)
 \delta^{(2)}(\q+\q') \, {\cal V}^2(\q) \,.
\eeq
Below, we list the $N=2$ amplitudes for the quark, while the corresponding expressions for the antiquark follow from symmetry.

\subsubsection{The contact terms}
The diagrams where the same propagator interacts twice with the medium are written in the contact limit as
\begin{multline}
\mathcal{M}_{(2)1}^a = \begin{minipage}{1.2cm} \includegraphics[width=\textwidth]{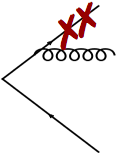} \end{minipage} \xrightarrow{\text{C.L.}} g^3\, \mathcal{M}_0(p,\bar p) \, t^{b}t^{b}t^a \, m_D^2 \int \frac{d^2 \q}{(2\pi)^2} \mathcal{V}^2(\q)\\ \times \int\!\!dx_1^+ \, n(x_1^+)  \frac{\kh \cdot \epsp}{\kh^2} e^{i k^- x_0^+} \left( 1- e^{i\frac{\kh^2}{2k^+}\Delta_{10}^+ }\right)  \,,
\end{multline}
where we have used that $\Theta(0)=1\big/2$. Similarly,
\begin{multline}
\mathcal{M}_{(2)2}^a = \begin{minipage}{1.2cm} \includegraphics[width=\textwidth]{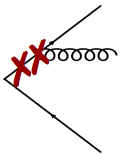} \end{minipage} \xrightarrow{\text{C.L.}} g^3\,\mathcal{M}_0(p,\bar p) \, t^{b}t^{b}t^a \, m_D^2 \int \frac{d^2 \q}{(2\pi)^2} \mathcal{V}^2(\q) \\ \times \int\!\!dx_1^+ \, n(x_1^+) \frac{\kh\cdot \epsp}{\kh^2} e^{i k^- x_0^+}  e^{i\frac{\kh^2}{2k^+}\Delta_{10}^+}  \,,
\end{multline}
\begin{multline}
\mathcal{M}_{(2)3}^a = \begin{minipage}{1.2cm} \includegraphics[width=\textwidth]{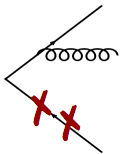} \end{minipage} \xrightarrow{\text{C.L.}} g^3\,\mathcal{M}_0(p,\bar p) \, t^at^{b}t^{b} \, m_D^2 \int \frac{d^2 \q}{(2\pi)^2}  \mathcal{V}^2(\q) \\ \times \int \!\!dx_1^+ \, n(x_1^+) \frac{\kh\cdot \epsp}{\kh^2} e^{i k^- x_0^+} \;,
\end{multline}
\begin{multline}
\mathcal{M}_{(2)4}^a = \begin{minipage}{1.2cm} \includegraphics[width=\textwidth]{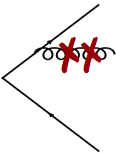} \end{minipage} \xrightarrow{\text{C.L.}} g^3\,\mathcal{M}_0(p,\bar p) \, [[t^a,t^{b}],t^{b}] \, m_D^2 \int \frac{d^2 \q}{(2\pi)^2} \mathcal{V}^2(\q) \\ \times \int\!\!dx_1^+ \, n(x_1^+)  \frac{\kh\cdot \epsp}{\kh^2} e^{i k^- x_0^+} \left( 1- e^{i\frac{\kh^2}{2k^+}\Delta_{10}^+ } \right)  \,,
\end{multline}

\subsubsection{Remaining double-interaction terms}
\begin{multline}
\mathcal{M}_{(2)5}^a = \begin{minipage}{1.2cm} \includegraphics[width=\textwidth]{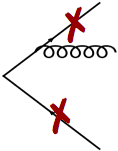} \end{minipage} \xrightarrow{\text{C.L.}} -2g^3\,\mathcal{M}_0(p,\bar p) \, t^{b}t^at^{b}\, m_D^2 \int  \frac{d^2 \q}{(2\pi)^2} \mathcal{V}^2(\q) \\ \times \int \!\! dx_1^+ n(x_1^+) \frac{\kh\cdot \epsp}{\kh^2} e^{i k^- x_0^+} e^{i \pip\cdot\q\, \Delta_{10}^+}\left(1-e^{i\frac{\kh^2}{2k^+}\Delta_{10}^+} \right) \,,
\end{multline}
\begin{multline} 
\mathcal{M}_{(2)6}^a = \begin{minipage}{1.2cm} \includegraphics[width=\textwidth]{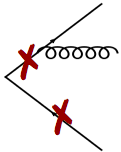} \end{minipage} \xrightarrow{\text{C.L.}} -2g^3\,\mathcal{M}_0(p,\bar p) \, t^at^{b}t^{b}\, m_D^2 \int \frac{d^2 \q}{(2\pi)^2}   \mathcal{V}^2(\q) \\  \times  \int\!\!dx_1^+ n(x_1^+) \frac{\kh\cdot \epsp}{\kh^2} e^{i k^- x_0^+}  e^{i \pip \cdot \q\,\Delta_{10}^+}e^{i\frac{\kh^2}{2k^+}\Delta_{10}^+}\,,
\end{multline}
\begin{multline}
\mathcal{M}_{(2)7}^a = \begin{minipage}{1.2cm} \includegraphics[width=\textwidth]{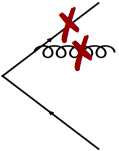} \end{minipage} \xrightarrow{\text{C.L.}} -2g^3\,\mathcal{M}_0(p,\bar p) \, t^{b}[t^a,t^b] \, m_D^2 \int  \frac{d^2 \q}{(2\pi)^2} \mathcal{V}^2(\q) \\ \times \int \!\!dx_1^+ n(x_1^+) \frac{(\Qh) \cdot \epsp}{(\Qh)^2} e^{i k^- x_0^+}  e^{i \frac{\kh^2}{2k^+}\Delta_{10}^+}\left(1-e^{-i\frac{(\Qh)^2}{2k^+}\Delta_{10}^+} \right) \,,
\end{multline}
\begin{multline}
\mathcal{M}_{(2)8}^a = \begin{minipage}{1.2cm} \includegraphics[width=\textwidth]{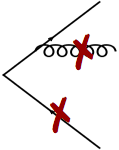} \end{minipage} \xrightarrow{\text{C.L.}} 2g^3\,\mathcal{M}_0(p,\bar p) \, [t^a,t^{b}]t^{b}\, m_D^2 \int \frac{d^2 \q}{(2\pi)^2} \mathcal{V}^2(\q)\\ \times \int \!\!dx_1^+ n(x_1^+)  \frac{(\Qh) \cdot \epsp}{(\Qh)^2} e^{i k^- x_0^+} e^{i\frac{\kh^2}{2k^+}\Delta_{10}^+} e^{i \pip \cdot \q \, \Delta_{10}^+} \left(1-e^{-i\Delta_{10}^+\frac{(\Qh)^2}{2k^+}} \right)  \,.
\end{multline}
Taking the square of the amplitudes in Sec.~\ref{sec:feynman-n1} and supplementing with the contact terms in Sec.~\ref{sec:feynman-n2} according to eq.~(\ref{eq:FeynCrossSection}), the total medium-induced spectrum written in detail reduces to the spectrum in eq.~(\ref{eq:SpectrumMed}) taken together with the contact terms in eq.~(\ref{eq:ContactTermDefinition}).  This proves the correspondence between the semi-classical calculation in Section~\ref{sec:vacuum} and \ref{sec:medium} and standard perturbation theory. One one hand, the simplicity of the results within the CYM formalism are striking compared to the Feynman diagram approach, where many diagrams cancel at the level of the cross section. On the other hand, the latter approach contains more information on the level of the amplitudes. E.g., one readily finds the color singlet projection of the antenna from the amplitudes in Sec.~\ref{sec:feynman-n1}, while this cannot be deduced from the sum of amplitudes given by the compact expression in eq.~(\ref{eq:amplitude}), obtained by solving the CYM equations.

\section{The cross section}
\label{sec:feynman-cs}
In the folllowing we put $x_0^+= 0$ and will not write explicitly the integrals over $x_1^+$ (simply $x^+$ below) or $\q$. We also drop a common pre-factor $g^4 m_D^2\big/(2(2\pi)^3)\mathcal{V}^2(\q)$ for all the following terms.

\subsection{The independent spectrum}
\label{sec:feynman-glv}
By this we denote diagrams where the emission happens from the same quark in both the amplitude and the complex conjugate. The relevant diagrams are given by
\begin{align}
\begin{minipage}{1.2cm}\includegraphics[width=\textwidth]{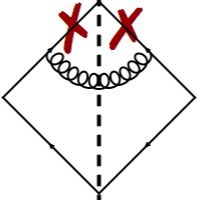}\end{minipage} &= 8\,\aabb\, \frac{1}{\kh^2} \left( 1- \cos \Omegaqzer
x^+\right)\,, \\
\begin{minipage}{1.2cm}\includegraphics[width=\textwidth]{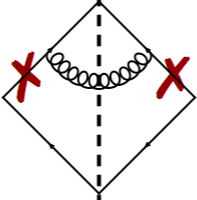}\end{minipage}  &=  4\,\aabb\, \frac{1}{\kh^2} \,, \\
\begin{minipage}{1.2cm}\includegraphics[width=\textwidth]{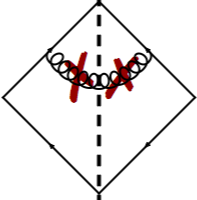} \end{minipage} &= 16\,\trdiff\,\frac{1}{(\Qh)^2} \left[ 1- \cos \Omegaq x^+\right] \,,
\end{align}
\begin{align}
\begin{minipage}{1.2cm}\includegraphics[width=\textwidth]{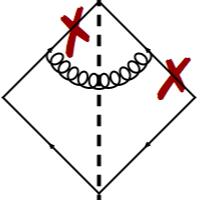} \end{minipage} +\text{c.c.} &= -8  \,\abab\,\frac{1}{\kh^2} \left( 1- \cos \Omegaqzer x^+\right),\\
\begin{minipage}{1.2cm}\includegraphics[width=\textwidth]{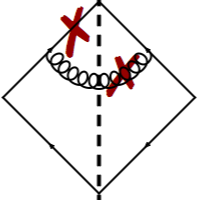}\end{minipage} +\text{c.c} &= -8  \,\trdiff\, \frac{\kh \cdot (\Qh) }{\kh^2 (\Qh)^2} \Bigg\{ 1- \cos \Omegaqzer x^+ -\cos \Omegaq x^+ \nn
&\qquad + \cos \left[\Omegaq - \Omegaqzer \right] x^+\Bigg\} \,, \\
\begin{minipage}{1.2cm}\includegraphics[width=\textwidth]{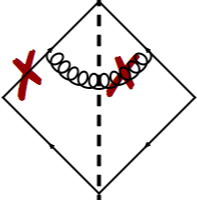}\end{minipage} +\text{c.c.} &= -8 \,\trdiff\, \frac{\kh \cdot (\Qh)}{\kh^2 (\Qh)^2} \left[ 1 -\cos \Omegaq x^+\right] \,, \\
\begin{minipage}{1.2cm} \includegraphics[width=\textwidth]{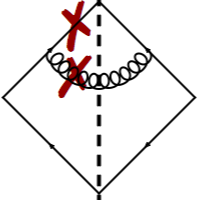} \end{minipage} +\text{c.c.} &= -8 \,\trdiff\, \frac{\kh \cdot (\Qh)}{\kh^2 (\Qh)^2} \left\{ \cos \Omegaqzer x^+ - \cos \left[\Omegaq - \Omegaqzer \right] x^+ \right\} \;,
\end{align}
where we have defined
\beq
\aabb &=& \frac{1}{N_c}\text{Tr} \big(t^a t^a t^bt^b \big)  \\
\abab &=& \frac{1}{N_c}\text{Tr} \big( t^a t^b t^a t^b \big) \,,
\eeq
and $\trdiff = \aabb - \abab = C_A C_F \big/2$. Additionally, we have the contribution to the cross section from the contact terms, given by
\begin{align}
\begin{minipage}{1.2cm}\includegraphics[width=\textwidth]{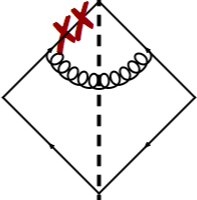}\end{minipage}  + \text{c.c.} &= - 4\,\aabb\, \frac{1}{\kh^2} \left( 1- \cos \Omegaqzer x^+\right) \,, \\
\begin{minipage}{1.2cm}\includegraphics[width=\textwidth]{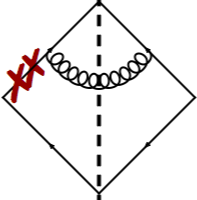}\end{minipage}  + \text{c.c.} &= -  4\,\aabb\, \frac{1}{\kh^2}  \cos \Omegaqzer x^+ \,,\\ 
\begin{minipage}{1.2cm}\includegraphics[width=\textwidth]{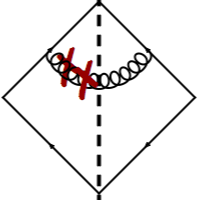}\end{minipage}  + \text{c.c.} &= -8 \,\trdiff\, \frac{1}{\kh^2} \left( 1- \cos \Omegaqzer x^+ \right)  \;.
\end{align}
The diagrams above correspond to the BDMPS-Z/GLV spectrum at first order in opacity off the quark and, summing up, yield 
\beq
\label{eq:diag_q}
8 C_A C_F \left[ 1-\cos\Omegaq x^+ \right] \frac{\kh \cdot \q}{\kh^2 (\Qh)^2} \,.
\eeq
The corresponding diagrams for the antiquark give
\beq
\label{eq:diag_qb}
8  C_A C_F \left[ 1-\cos\Omegaqb x^+ \right]  \frac{\khb \cdot \q}{\khb^2 (\Qhb)^2}  \,.
\eeq
What remains to prove is that new diagrams involving emission off the quark and re-scattering of the antiquark cancel, and vice versa. First and foremost, it is easy to show that
\beq
\begin{minipage}{1.2cm} \includegraphics[width=\textwidth]{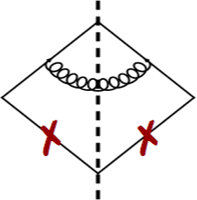} \end{minipage} + \begin{minipage}{1.2cm}\includegraphics[width=\textwidth]{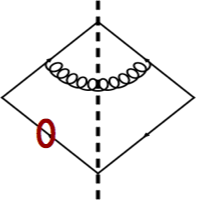}\end{minipage} + \mbox{c.c.} =0 \;.
\eeq
Additionally we have the following identities
\begin{align}
\begin{minipage}{1.2cm}\includegraphics[width=\textwidth]{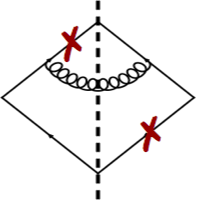} \end{minipage} + \begin{minipage}{1.2cm}\includegraphics[width=\textwidth]{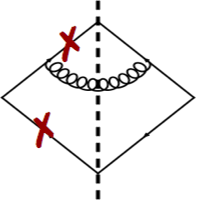} \end{minipage} +\mbox{c.c.} &= 0 \\
\begin{minipage}{1.2cm}\includegraphics[width=\textwidth]{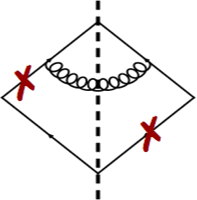} \end{minipage} + \begin{minipage}{1.2cm}\includegraphics[width=\textwidth]{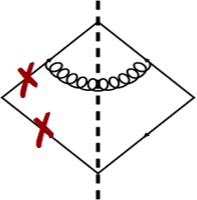} \end{minipage} +\mbox{c.c.} &= 0 \\
\begin{minipage}{1.2cm}\includegraphics[width=\textwidth]{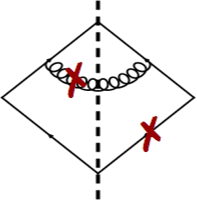} \end{minipage} + \begin{minipage}{1.2cm}\includegraphics[width=\textwidth]{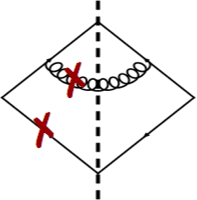} \end{minipage} +\mbox{c.c.} &= 0
\end{align}
The same is true for the antiquark.

\subsection{The ``gluon interference" terms}
\label{sec:feynman-gluon}
First we calculate the diagram where the gluon interacts with the medium in both amplitude and complex conjugate amplitude. Only one diagram give rise to this specific term, namely
\begin{multline}
\label{eq:interf_gg}
\begin{minipage}{1.2cm}\includegraphics[width=\textwidth]{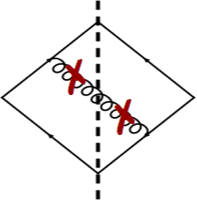}\end{minipage} + \mbox{c.c.} = -16 \,\trdiff\,\frac{(\Qh) \cdot (\Qhb)}{(\Qh)^2 (\Qhb)^2} \bigg[ 1+ \cos \Omegaqqb x^+ \\ - \cos \Omegaq x^+ - \cos\Omegaqb x^+ \bigg] \,,
\end{multline}
where we have defined
\beq
\Omegaqqb \equiv \frac{\left( \kh + \khb-2\q\right) \cdot \pip}{2} \,,
\eeq
to shorten the notation. Furthermore, there are 5 interference diagrams for the where the gluon is emitted from the quark and interacts only once with the medium. Four of them are given by the identities
\begin{align}
\begin{minipage}{1.2cm} \includegraphics[width=\textwidth]{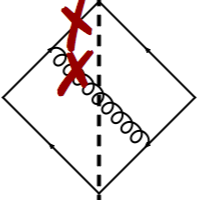} \end{minipage} + \begin{minipage}{1.2cm} \includegraphics[width=\textwidth]{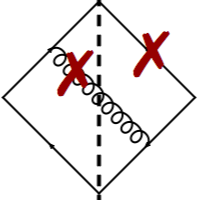} \end{minipage} +\text{c.c.} &=0 \;, \\
 \begin{minipage}{1.2cm} \includegraphics[width=\textwidth]{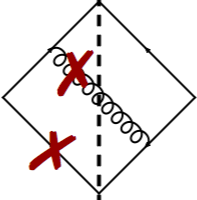} \end{minipage}  + \begin{minipage}{1.2cm} \includegraphics[width=\textwidth]{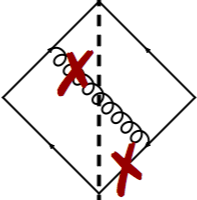} \end{minipage} +\text{c.c.} &= 0 \;,
\end{align}
while the only remaining term read
\beq
\label{eq:interf_gq}
\begin{minipage}{1.2cm} \includegraphics[width=\textwidth]{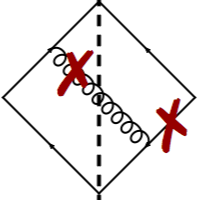} \end{minipage} +\text{c.c.} = 8\, C_A C_F \frac{\khb \cdot (\Qh)}{\khb^2 (\Qh)^2} \left[\cos \Omegaqqb x^+ - \cos \Omegaqb x^+ \right] \,,
\eeq
and analogously for the antiquark.

\subsection{The ``quark bremsstrahlung" interference}
\label{sec:feynman-brems}
By definition, these are the diagrams where the gluon does not interact with the medium. They are given by
\begin{align}
\begin{minipage}{1.2cm}\includegraphics[width=\textwidth]{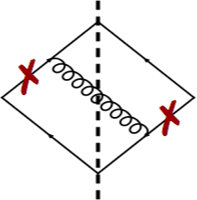} \end{minipage} + \mbox{c.c.} &= 8 \,\abab\, \frac{\kh \cdot \khb}{\kh^2 \khb^2} \cos \Omegaqqb x^+ \,,\\
\begin{minipage}{1.2cm}\includegraphics[width=\textwidth]{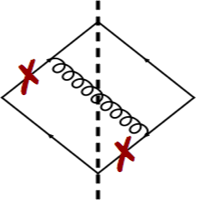}\end{minipage} + \mbox{c.c.} &= -8 \,\aabb\, \frac{\kh \cdot \khb}{\kh^2 \khb^2} \left[ \cos \Omegaqqb x^+  - \cos \left(\Omegaqqb+ \Omegaqbzer \right)x^+ \right] \,, \\
\begin{minipage}{1.2cm}\includegraphics[width=\textwidth]{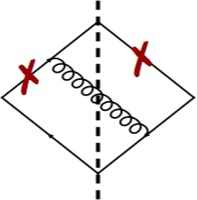} \end{minipage}+ \mbox{c.c.} &= -8\,\abab\, \frac{\kh\cdot \khb}{\kh^2 \khb^2} \cos \Omegaqzer x^+ \,,
\end{align}
\begin{align}
\begin{minipage}{1.2cm}\includegraphics[width=\textwidth]{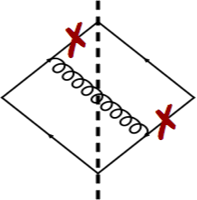} \end{minipage}+ \mbox{c.c.} &= -8 \,\aabb\,\frac{\kh \cdot \khb}{\kh^2 \khb^2} \left[ \cos \Omegaqqb x^+ - \cos \left( \Omegaqqb - \Omegaqzer \right)x^+ \right] \,, \\
\begin{minipage}{1.2cm}\includegraphics[width=\textwidth]{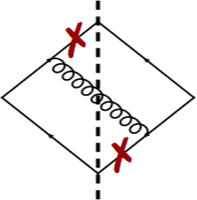} \end{minipage}+ \mbox{c.c.} &= 8 \,\abab\, \frac{\kh \cdot \khb}{\kh^2 \khb^2} \bigg[ \cos \Omegaqqb x^+ - \cos \left(\Omegaqqb + \Omegaqbzer \right)x^+ \nn
&\qquad - \cos \left(\Omegaqqb - \Omegaqzer \right)x^+ +\cos \left(\Omegaqqb - \Omegaqzer + \Omegaqbzer \right)x^+\bigg] \,,\\
\begin{minipage}{1.2cm}\includegraphics[width=\textwidth]{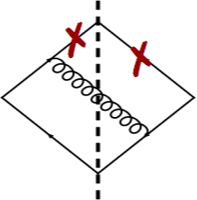} \end{minipage}+ \mbox{c.c.} &= 8 \,\aabb\, \frac{\kh \cdot \khb}{\kh^2 \khb^2} \left( \cos \Omegaqzer x^+ -1 \right) \,,
\end{align}
\begin{align}
\begin{minipage}{1.2cm}\includegraphics[width=\textwidth]{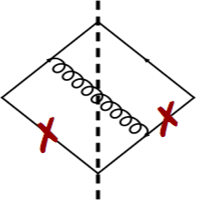} \end{minipage}+ \mbox{c.c.} &= -8 \,\abab\, \frac{\kh \cdot \khb}{\kh^2 \khb^2} \cos \Omegaqzer x^+ \,, \\
\begin{minipage}{1.2cm}\includegraphics[width=\textwidth]{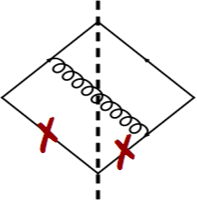} \end{minipage}+ \mbox{c.c.} &= 8 \,\aabb\,\frac{\kh \cdot \khb}{\kh^2 \khb^2} \left(\cos \Omegaqbzer x^+ - 1\right) \, \\
\begin{minipage}{1.2cm}\includegraphics[width=\textwidth]{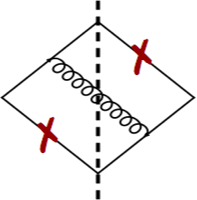} \end{minipage}+ \mbox{c.c.} &= 8 \,\abab\,\frac{\kh \cdot \khb}{\kh^2 \khb^2} \cos \pip \cdot \q x^+ \,,
\end{align}
where $\pip$ is defined in eq.~(\ref{eq:deltan}).

Additionally, we have the double interaction diagrams. First, the contact terms read as follows
\begin{align}
\begin{minipage}{1.2cm}\includegraphics[width=\textwidth]{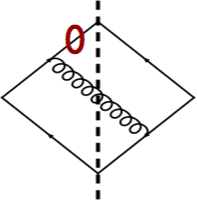}\end{minipage} + \mbox{c.c.} &= 4\aabb\, \frac{\kh \cdot \khb}{\kh^2 \khb^2} \left(1- \cos \Omegaqzer x^+\right) \,,\\
\begin{minipage}{1.2cm}\includegraphics[width=\textwidth]{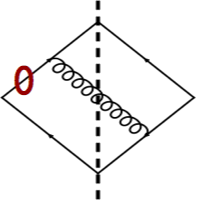}\end{minipage} + \mbox{c.c.} &= 4\aabb\, \frac{\kh \cdot \khb}{\kh^2 \khb^2} \cos \Omegaqzer x^+ \,,
\end{align}
\begin{align}
\begin{minipage}{1.2cm}\includegraphics[width=\textwidth]{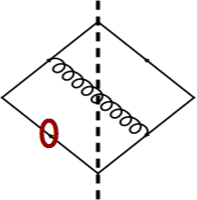}\end{minipage} + \mbox{c.c.} &= 4\aabb\, \frac{\kh \cdot \khb}{\kh^2 \khb^2} \,, \\
\begin{minipage}{1.2cm}\includegraphics[width=\textwidth]{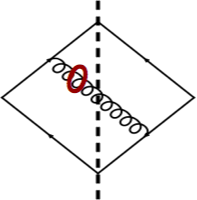}\end{minipage} + \mbox{c.c.} &= - 8 \,\trdiff\,\frac{\kh \cdot \khb}{\kh^2 \khb^2} \left( \cos \Omegaqzer x^+ -1\right) \,,
\end{align}
and we also have to include
\begin{align}
\begin{minipage}{1.2cm}\includegraphics[width=\textwidth]{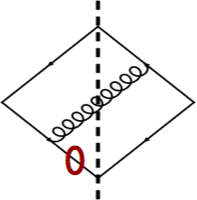} \end{minipage}+ \mbox{c.c.} &= 4\aabb\,\frac{\kh \cdot \khb}{\kh^2 \khb^2} \left(1- \cos \Omegaqbzer x^+\right)\,, \\
\begin{minipage}{1.2cm}\includegraphics[width=\textwidth]{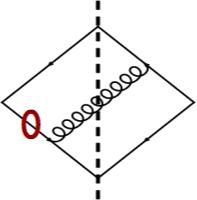}\end{minipage} + \mbox{c.c.} &=4\aabb \,\frac{\kh \cdot \khb}{\kh^2 \khb^2} \cos \Omegaqbzer x^+ \,, \\
\begin{minipage}{1.2cm}\includegraphics[width=\textwidth]{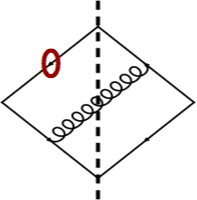}\end{minipage} + \mbox{c.c.} &=4 \aabb\, \frac{\kh \cdot \khb}{\kh^2 \khb^2}\,,\\
\begin{minipage}{1.2cm}\includegraphics[width=\textwidth]{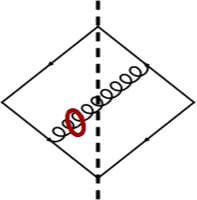}\end{minipage} + \mbox{c.c.} &=-8\, \trdiff\, \frac{\kh \cdot \khb}{\kh^2 \khb^2} \left( \cos \Omegaqbzer x^+ -1\right) \,.
\end{align}
Finally, we have the remaining double interaction terms
\begin{align}
\begin{minipage}{1.2cm}\includegraphics[width=\textwidth]{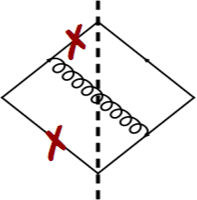}\end{minipage} + \mbox{c.c.} &= -8\,\abab\, \frac{\kh \cdot \khb}{\kh^2 \khb^2} \left[ \cos \pip \cdot \q x^+ - \cos \left( \Omegaqzer + \pip \cdot \q\right)x^+\right] \,,\\
\begin{minipage}{1.2cm}\includegraphics[width=\textwidth]{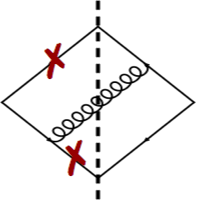}\end{minipage} + \mbox{c.c.} &= -8 \,\abab\,\frac{\kh \cdot \khb}{\kh^2 \khb^2} \left[ \cos \pip\cdot \q x^+ - \cos \left( \Omegaqbzer - \pip \cdot \q\right)x^+\right] \,, \\
\begin{minipage}{1.2cm}\includegraphics[width=\textwidth]{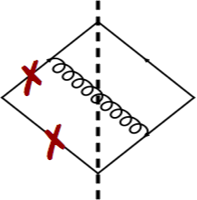} \end{minipage}+ \mbox{c.c.} &= -8 \,\aabb\, \frac{\kh \cdot \khb}{\kh^2 \khb^2} \cos \left( \Omegaqzer + \pip\cdot \q\right)x^+ \,,\\
\begin{minipage}{1.2cm}\includegraphics[width=\textwidth]{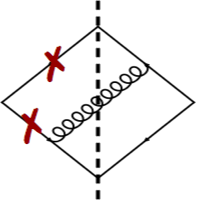}\end{minipage} + \mbox{c.c.} &= -8 \,\aabb\,\frac{\kh \cdot \khb}{\kh^2 \khb^2} \cos \left( \Omegaqbzer - \pip \cdot \q\right)x^+ \,.
\end{align}
Summing these terms we obtain 
\beq
\label{eq:interf_brems}
8\, C_A C_F\left(1- \cos \Omegaqqb x^+ \right)\frac{\kh \cdot \khb}{\kh^2 \khb^2} \;.
\eeq

The sum of eqs. (\ref{eq:diag_q}), (\ref{eq:diag_qb}), (\ref{eq:interf_gg}), (\ref{eq:interf_gq}) (and the one from the anti-quark) and (\ref{eq:interf_brems}) gives
the final result in eq. (\ref{eq:SpectrumMed}), which is used throughout the paper.

\end{document}